\documentclass[preprint,secnumarabic,amssymb,amsmath, nobibnotes, aps, prfluids]{revtex4-2}

\newcommand{\env}[1]{\texttt{#1}}
\usepackage{graphicx}
\usepackage{hyperref}
\usepackage{xcolor}

\def\change#1{{{#1}}}

\begin{document}

\title{Physical invariance in neural networks for subgrid-scale scalar flux modeling}%

\author{Hugo Frezat}
\email{hugo.frezat@univ-grenoble-alpes.fr}
\affiliation{University Grenoble Alpes, CNRS UMR LEGI, Grenoble, France}
\author{Guillaume Balarac}
\affiliation{University Grenoble Alpes, CNRS UMR LEGI, Grenoble, France}
\affiliation{Institut Universitaire de France (IUF), Paris, France}
\author{Julien \surname{Le Sommer}}
\affiliation{University Grenoble Alpes, CNRS UMR IGE, Grenoble, France}
\author{Ronan Fablet}
\affiliation{IMT Atlantique, CNRS UMR Lab-STICC, Brest, France}
\author{Redouane Lguensat}
\affiliation{Laboratoire des Sciences du Climat et de l'Environnement (LSCE), IPSL/CEA, Gif Sur Yvette, France}
\affiliation{\change{LOCEAN-IPSL, Sorbonne Universit\'e, Institut Pierre Simon Laplace, Paris, France}}

\date{\today}%

\begin{abstract}
In this paper we present a new strategy to model the subgrid-scale scalar flux in a three-dimensional turbulent incompressible flow using physics-informed neural networks (NNs). When trained from direct numerical simulation (DNS) data, state-of-the-art neural networks, such as convolutional neural networks, may not preserve well known physical priors, which may in turn question their application to real case-studies.
To address this issue, we investigate hard and soft constraints into the model based on classical transformation invariances and symmetries derived from physical laws.
From simulation-based experiments, we show that the proposed transformation-invariant NN model outperforms both purely data-driven ones as well as parametric state-of-the-art subgrid-scale models.
The considered invariances are regarded as regularizers on physical metrics during the a priori evaluation and constrain the distribution tails of the predicted subgrid-scale term to be closer to the DNS.
They also increase the stability and performance of the model when used as a surrogate during a large-eddy simulation. Moreover, the transformation-invariant NN is shown to generalize to \change{regimes} that have not been seen during the training phase.
\end{abstract}

\maketitle

\section{Introduction}
The transport of a scalar quantity within a flow arises as a key issue in many applications, ranging from atmosphere and ocean physics to fluid dynamics for industrial purposes. While the scalar transport equations are known, solving this problem is still a difficult numerical challenge due to the large range of motion scales encountered in turbulent flows. As such, the direct numerical simulation (DNS) of realistic applications is not yet possible and reduced-order frameworks such as the large-eddy simulation (LES) have gained interest. The principle of the LES is to only compute the large scales of the flow and take into account the interaction of the smallest scales on the large scale dynamics through a subgrid-scale (SGS) model.

Following the recent advances in deep learning for turbulence modeling \cite{duraisamy2019turbulence, wang2020towards, portwood2019turbulence, mohan2020embedding}, neural networks (NNs) are expected to be good candidates for formulating subgrid closures and offer a promising alternative to algebraic models. Initial works in this direction were initiated by the modeling of the SGS momentum closure in incompressible \cite{gamahara2017searching, xie2020modeling}, two dimensional \cite{maulik2019subgrid} or compressible \cite{xie2019artificial} turbulence, among others. Due to their computational advantages, recent studies have investigated convolutional neural networks (CNNs) \cite{beck2019deep,pawar2020priori}. They have shown that CNNs outperform algebraic models in specific applications, e.g. in the estimation of the subgrid-scale reaction rates \cite{lapeyre2019training} or the subgrid parametrization of ocean dynamics \cite{bolton2019applications}. However, these models do lack of embedded physical laws and are often thought as "black box" in which our comprehension is limited. This has two major implications for the model; first, the SGS model is only able to predict a narrow range of flows that have been used during training, and suffer from rather limited generalization (or extrapolation) capabilities. Then, and most importantly, the SGS model will situationally predict unrealistic SGS terms that do not follow the expected behavior from well known physical laws. Explicit embedding of known invariance has already shown great success \cite{ling2016reynolds, ling2016machine} when applied to Galilean invariance in Reynolds Averaged Navier Stokes (RANS) simulation. 

This paper addresses physical invariances in neural networks to improve the interpretability and generalization properties of SGS NN models. Whereas physical priors may be embedded implicitly through a custom training loss function \cite{bode2021using}, we rather explore how to explicitly embed expected physical invariances within a NN architecture. Through a case study on different \change{regimes} of scalar mixing in three-dimensional homogeneous isotropic turbulence, we show that the resulting NN framework significantly outperforms both state-of-the-art algebraic and NN models.

The paper is organized as follows: In Sec. II we introduce the SGS closure term that arises from the filtered transport equation and the modeling problem. Sec. III describes the numerical setup and baseline models used in our experiments along with different metrics for the evaluation. Demonstrating why and how physical invariances can be embedded into a NN is presented in Sec. IV. Finally, the evaluation a priori and a posteriori are presented \change{for} three different \change{regimes} in Sec. V.

\section{SGS scalar modeling}
In this section we introduce the incompressible Navier-Stokes equations that govern the flow dynamics,
\begin{equation}
    \frac{\partial \mathbf{u}}{\partial t} + (\mathbf{u} \cdot \nabla)\mathbf{u} = -\frac{1}{\rho} \nabla p + \nu \nabla^{2}\mathbf{u} + F_{\mathbf{u}}, \hspace{10mm} \nabla \cdot \mathbf{u} = 0,
    \label{eq:navierstokes}
\end{equation}
with velocity $\mathbf{u} = (u_{x}, u_{y}, u_{z})$, density $\rho$, pressure $p$, kinematic viscosity $\nu$ and $F_{\mathbf{u}}$ an external forcing. We define the advection-diffusion of a scalar $\Phi$ by its Schmidt number, molecular diffusivity $\kappa = \nu / Sc$ and source term $F_{\Phi}$
\begin{equation}
    \frac{\partial \Phi}{\partial t} + (\mathbf{u} \cdot \nabla) \Phi = \nabla \cdot (\kappa \nabla \Phi) + F_{\Phi}.
    \label{eq:advection}
\end{equation}
The idea behind LES is to apply a scale separation that filters out the smallest scales of the scalar field $\Phi$ such that
\begin{equation}
    \overline{\Phi(x)} = \int_{V} \Phi(x_{f})\,G(x_{f} - x)\,dx_{f},
    \label{eq:filter}
\end{equation}
where the kernel $G$ is a spectral cut-off kernel defined in spectral space by the Heaviside step function $H$
\begin{equation}
    \hat{G}(k) = H \left(\pi / \Delta - |k| \right),
\end{equation}
where $\Delta$ is the filter width. Applying the filter \eqref{eq:filter} on \eqref{eq:advection} yields the filtered advection-diffusion equation
\begin{equation}
    \frac{\partial \overline{\Phi}}{\partial t} + (\overline{\mathbf{u}} \cdot \nabla) \overline{\Phi} = \nabla \cdot (\kappa \nabla \overline{\Phi}) + \overline{F}_{\Phi} + \nabla \cdot (\overline{\mathbf{u}\,\Phi} - \overline{\mathbf{u}}\,\overline{\Phi})
    \label{eq:filtered_advection}
\end{equation}
in which the residual flux is defined exactly as,
\begin{equation}
    \mathbf{s} \equiv \overline{\mathbf{u}\,\Phi} - \overline{\mathbf{u}}\,\overline{\Phi}.
\end{equation}
Thus, the unknown SGS term that appears in \eqref{eq:filtered_advection} is the divergence of the residual flux, $\nabla \cdot \mathbf{s}$.
Building a model that closes such equations is usually done following 
\textit{functional} or \textit{structural} approaches \cite{sagaut2006large}. The functional strategy is based on physical considerations of the residual flux and not on the SGS term directly. Most functional models are based on the concept of eddy viscosity (or eddy diffusivity for a scalar) as in the Smagorinsky model \cite{smagorinsky1963general} and have been widely used in practice. While being limited in the range of dynamics that can be reproduced, these models have the advantage to be stable in \textit{a posteriori} evaluations, i.e., when studying their time evolution in a LES. 
The structural strategy on the other side aims at obtaining the best local approximation of the SGS term using the known small-scale structures of the flow. Structural models are built on a mathematical basis, often using a Taylor series expansion \cite{clark1979evaluation} or scale-similarity assumptions \cite{bardina1980improved}. These models yield the best performance in \textit{a priori} evaluations but are known to be unstable in long-term evolution due to their incorrect predictions of the subgrid local dissipation. 
Many improvements on the functional strategy have been explored, e.g., adding a dynamic coefficient \cite{germano1991dynamic, lilly1992proposed} extends the capabilities of the model to a wider range of \change{regimes}, while regularization schemes for the structural strategy give rise to mixed models \cite{balarac2013dynamic} using combinations of ideas from functional modeling.

With the emergence of deep learning \cite{lecun2015deep},
a renewed interest has been given to learning-based framework, especially NNs, as an alternative to algebraic schemes; see Refs. \cite{vollant2017subgrid, portwood2021interpreting}. NN models result in parametric SGS models, whose parameters are trained to minimize some error over a training set. However, to our knowledge, as illustrated in Sec. IV, such NN models do not convey expected physical properties (e.g., invariance properties), which greatly impact their actual range of applicability.

\section{Numerical setting}
NN SGS models require the generation of a training dataset, usually coming from DNS, from which model parameters are learnt. This dataset should contain a range of dynamics that are expected to be reproduced by the SGS model.
When dealing with SGS scalar turbulence modeling, we follow the hypothesis given by Batchelor \cite{batchelor1959small}, which states that the small scales of a scalar quantity in a turbulent flow at high Reynolds number are statistically similar. Thus, the SGS dataset shall comprise all the scales up to the limit between the inertial subrange and the energy containing range. To be precise, in the scalar-variance spectrum \cite{obukhov1970structure, corrsin1951spectrum} given by
\begin{equation}
    E_{\Phi}(k) = C_{\Phi} \chi \epsilon^{-1/3}k^{-5/3},
\end{equation}
where $C_{\Phi}$ is the Oboukou-Corrsin constant, $\chi$ and $\epsilon$ are the scalar-variance and kinetic-energy dissipation rate respectively in spectral space at wavenumber $k$. We want to model the part of the spectrum that follows the $k^{-5/3}$ exponential form, i.e., in the inertial-convective subrange, which is known to be statistically universal within a considered turbulence \change{regime}.

\subsection{Data generation}

\begin{figure*}[t]
  \centering
  \begin{minipage}[c]{.49\textwidth}
    \centering
    \includegraphics[width=1.\linewidth]{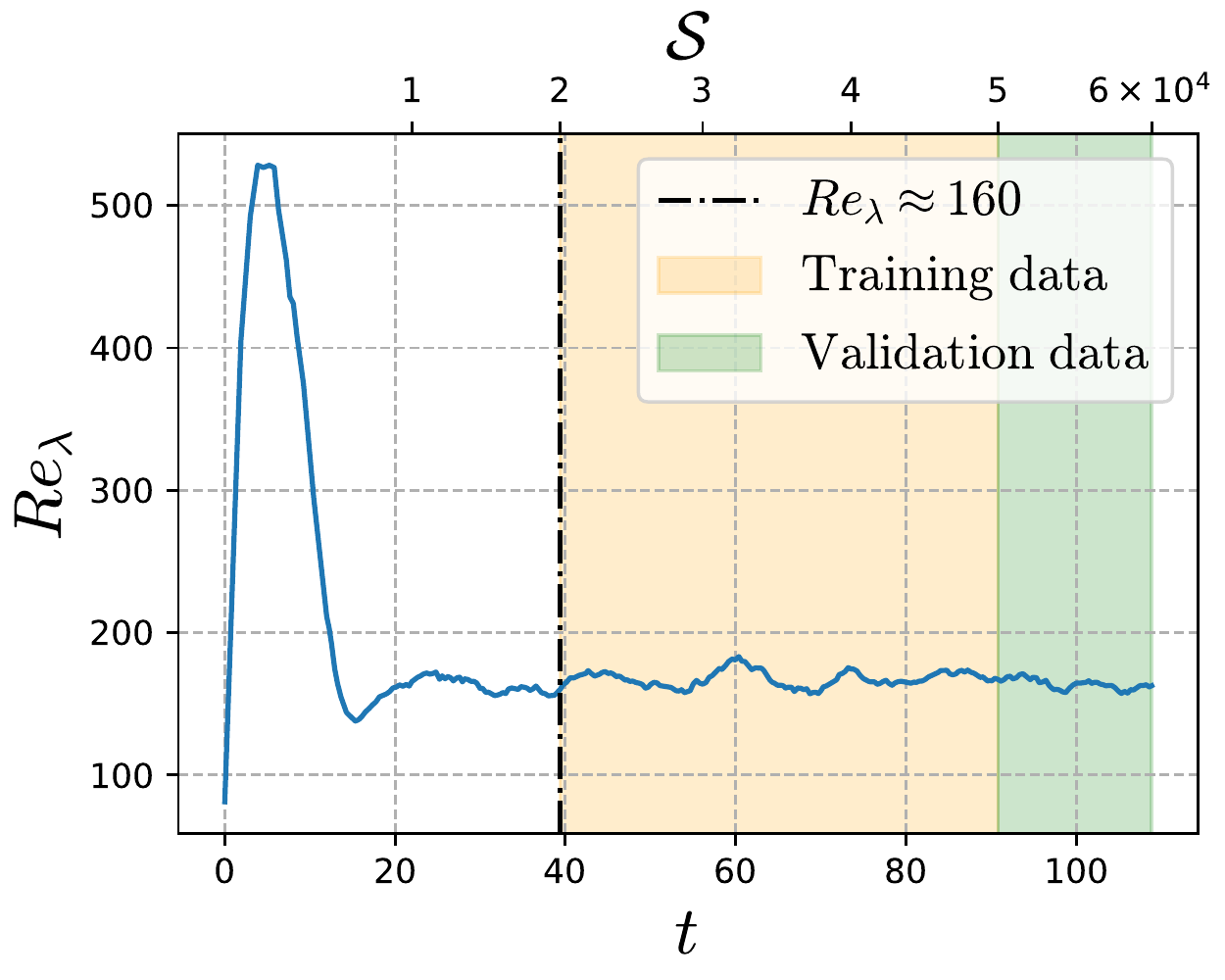}
  \end{minipage}
  \begin{minipage}[c]{.49\textwidth}
    \centering
    \includegraphics[width=1.\linewidth]{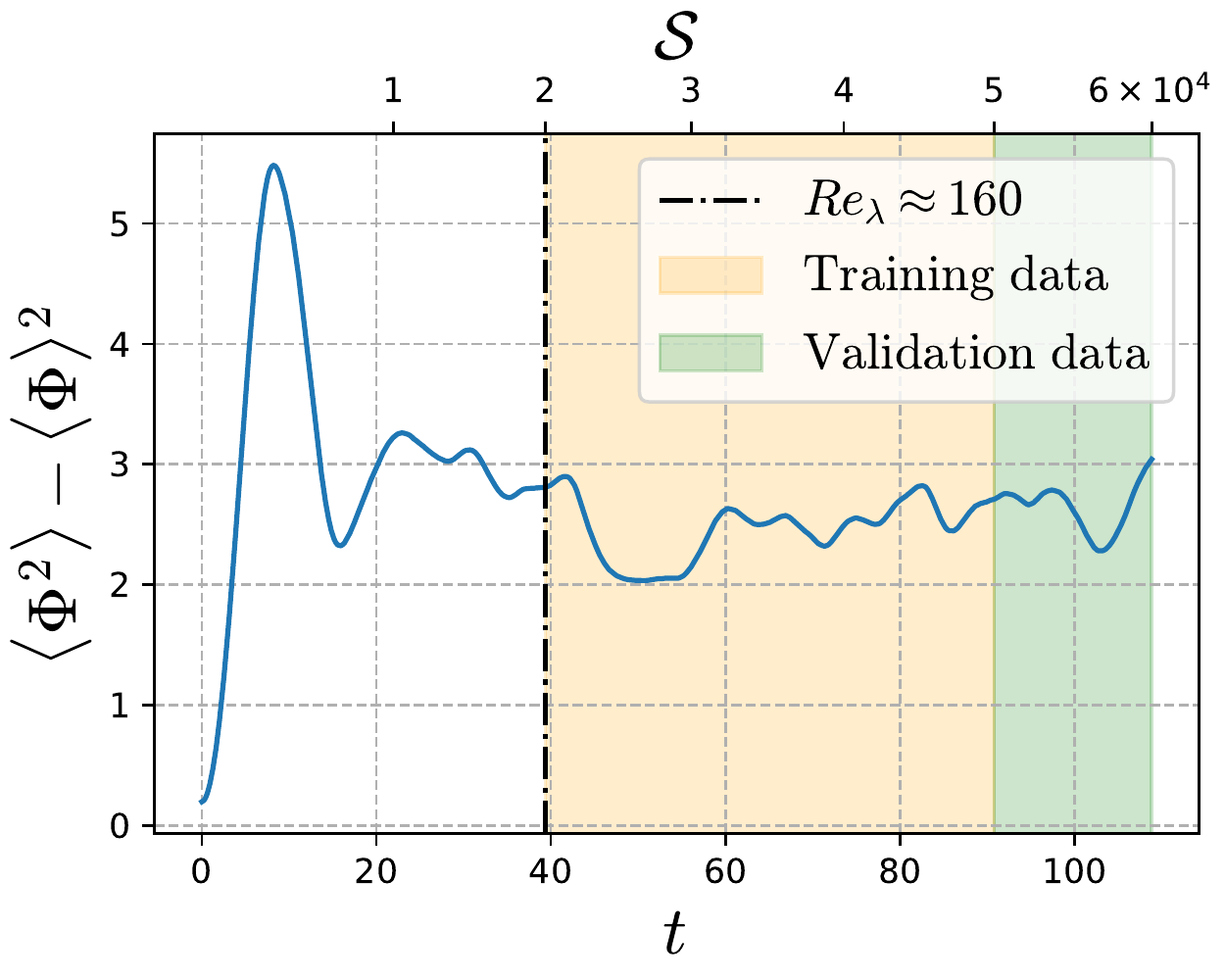}
  \end{minipage}
  \caption{Evolution of the Reynolds number on the Taylor microscale (left) and scalar variance (right). The flow is considered in an established turbulence regime at iteration $\mathcal{S} = 20000$. The first $50$ samples are used for training and the remaining $40$ samples are used for validation.
  \label{fig:train_data}
  }
\end{figure*}
\begin{figure*}[t]
  \centering
  \begin{minipage}[c]{.49\textwidth}
    \centering
    \includegraphics[width=1.\linewidth]{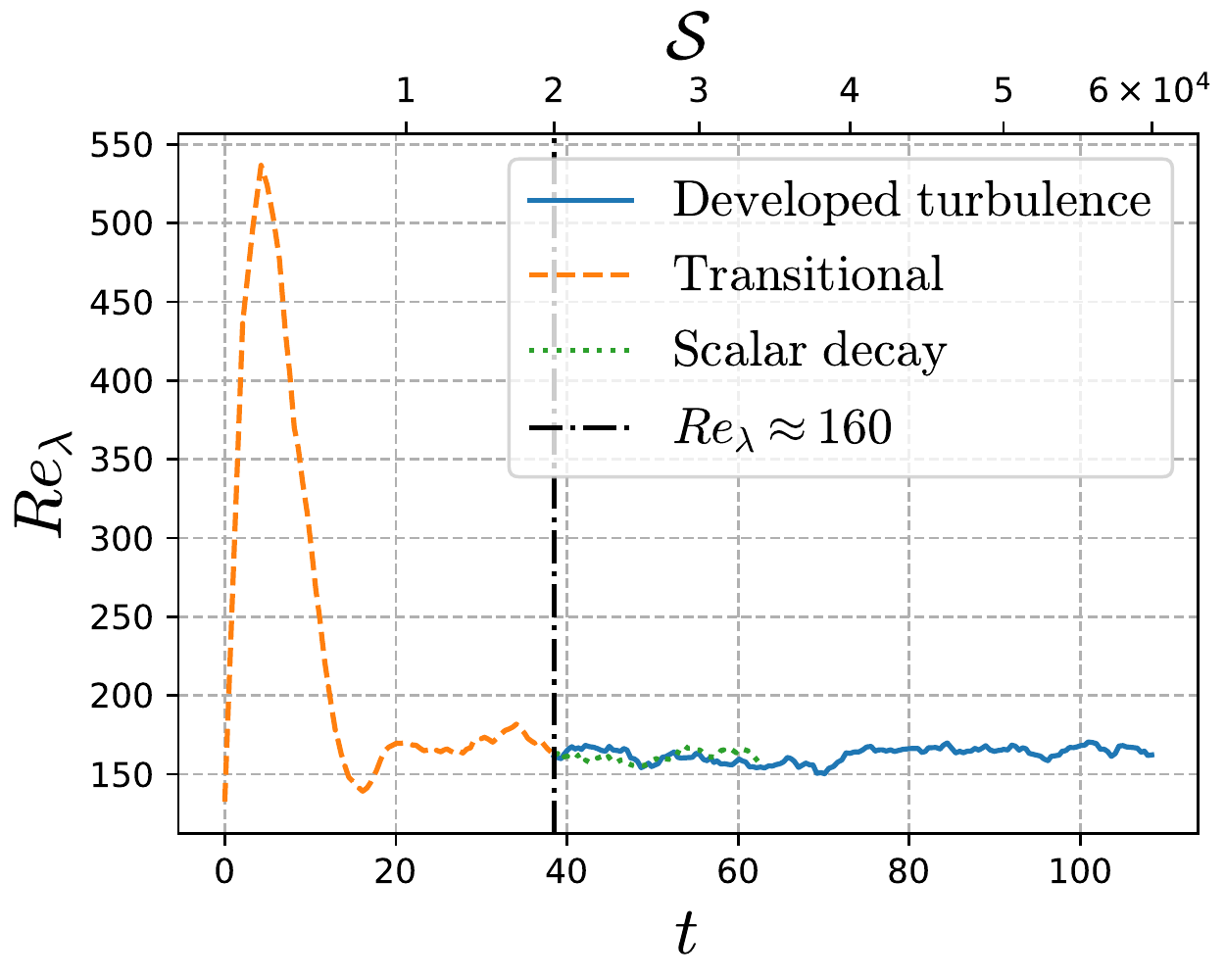}
  \end{minipage}
  \begin{minipage}[c]{.49\textwidth}
    \centering
    \includegraphics[width=1.\linewidth]{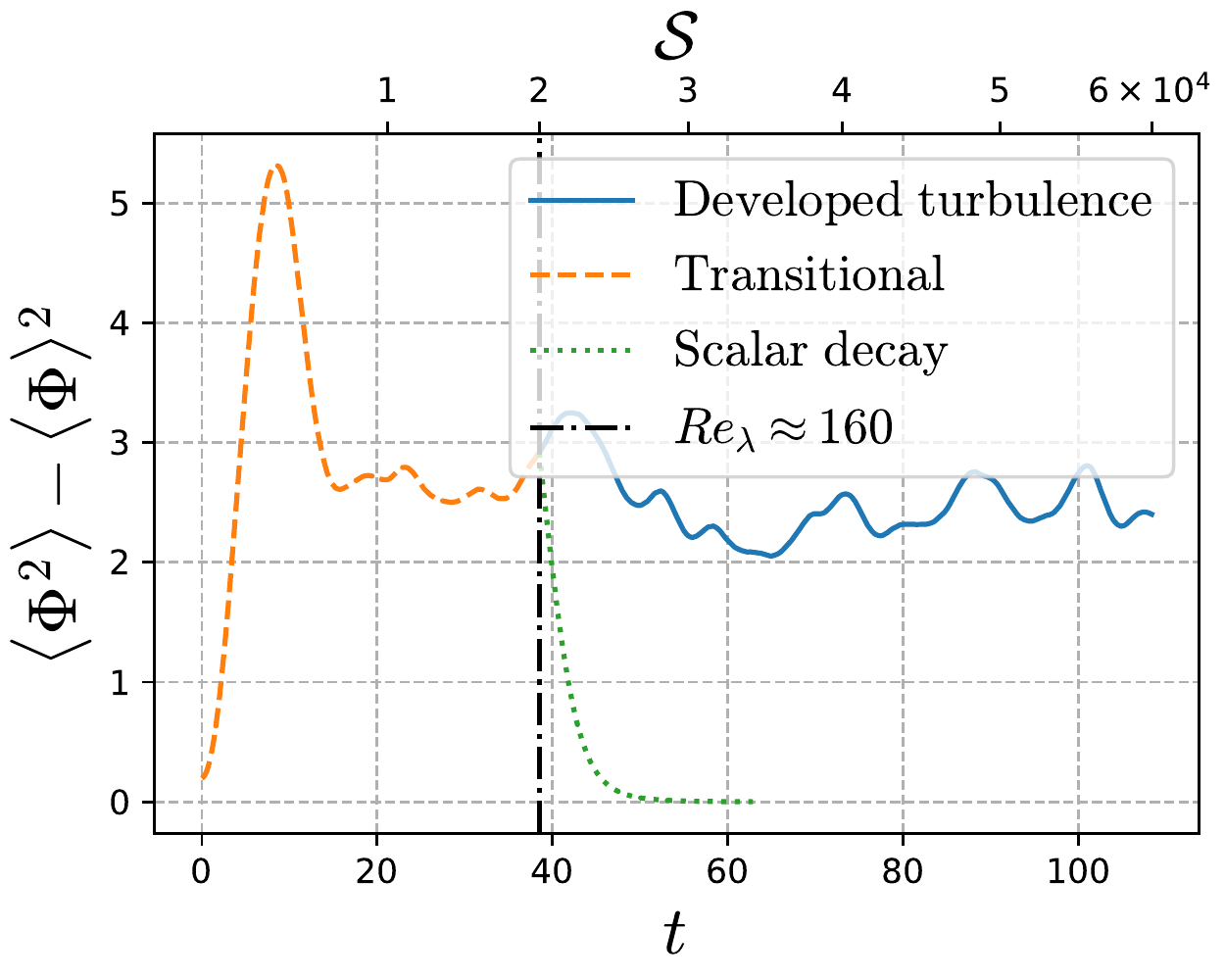}
  \end{minipage}
  \caption{Evolution of the Reynolds number on the Taylor microscale (left) and scalar variance (right) in the different regimes (developed turbulence, transitional and scalar decay) for the testing data.
  \label{fig:tests_data}
  }
\end{figure*}

Data used in this work are generated from a DNS of a forced homogeneous isotropic turbulence. A classical pseudospectral code with second-order explicit Runge-Kutta time advancement is used to solve equations \eqref{eq:navierstokes} and \eqref{eq:advection} in a triply periodic domain $[0, 2\pi)^3$. 
The size of the computational domain is larger than four times the integral length scale to ensure that the largest flow structures are not affected, and the domain is discretized using $512^3$ grid points. The simulation parameters are chosen such that $k_{\mathrm{max}}\eta > 1.5$ and $k_{\mathrm{max}}\eta_B > 1.5$, where $k_{\rm max}$ is the maximum wavenumber in the domain, and $\eta$ and $\eta_B$ are the Kolmogorov and Batchelor scales, respectively. The Reynolds number based on the Taylor microscale is around 160 and the molecular Schmidt number is set to 0.7. 
The initial random  velocity field is generated in Fourier space as a solenoidal isotropic field with random phases and a prescribed energy spectrum as in Ref. \cite{rosales2005linear}, for example. The initial scalar field is a large scale random field, generated according to the procedure proposed in Ref. \cite{eswaran1988examination}. Statistical stationarity is obtained using a random forcing located at low wavenumbers, $k \in [2, 3]$, for both velocity \cite{alvelius1999random} and scalar \cite{da2007analysis} fields. Applying the scalar spectral forcing only on the smallest wavenumbers allows us to contain its effect to the resolved scales of the simulation. Note that the code has been intensively used for physical analysis of turbulent flows \cite{gorbunova2020analysis}, numerical scheme development \cite{lagaert2014hybrid} and SGS models analysis \cite{balarac2008development, vollant2017subgrid}.

\begin{figure*}[t]
  \centering
  \includegraphics[width=1.0\textwidth]{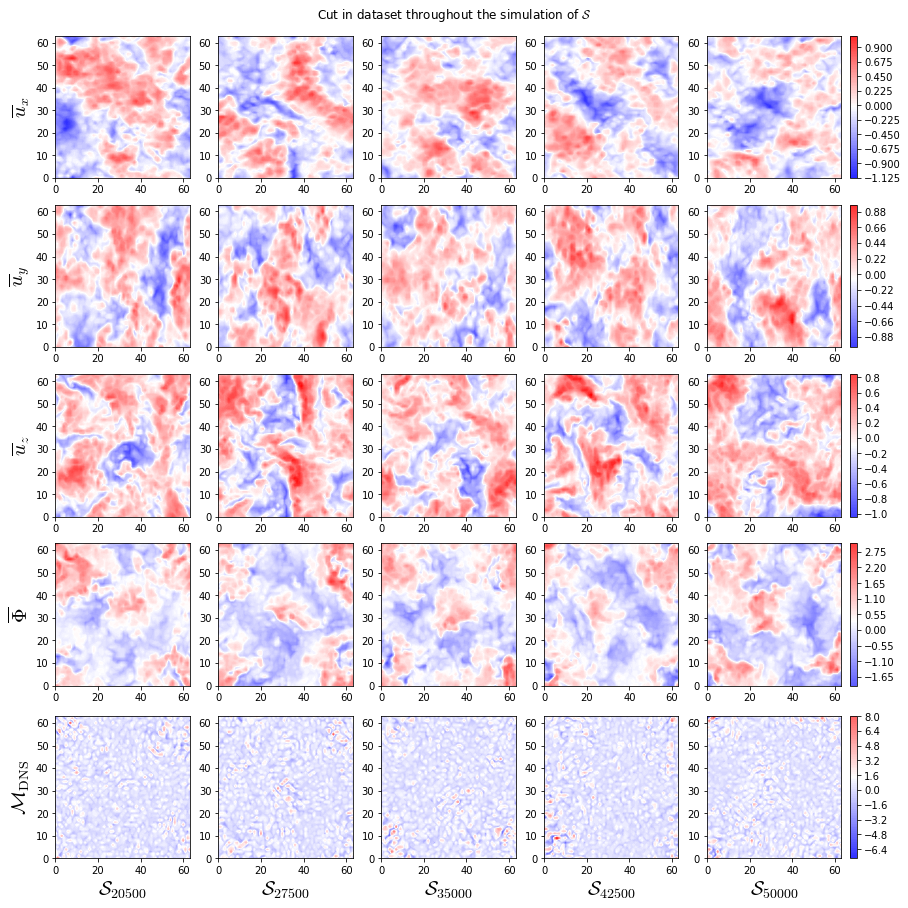}
  \caption{Data slices extracted during the simulation. $\overline{\mathbf{u}}$ are the velocities in each directions, $\overline{\Phi}$ is the scalar quantity and $\mathcal{M}_{\mathrm{DNS}}$ is the SGS term. The simulation was run with $N = 512$ grid points and filtered at $N = 64$. The flow reaches a statistically developed turbulence state approximately at iteration $\mathcal{S} = 20000$.
  \label{fig:dataset}
  }
\end{figure*}

Thus, we extract the filtered SGS term $\mathcal{M}_{\mathrm{DNS}}$ and nonresidual input quantities that appears in \eqref{eq:filtered_advection},
\begin{equation}
    \mathcal{M}_{\mathrm{DNS}}(\mathbf{i}) = \nabla \cdot \mathbf{s}, \hspace{10mm} \mathbf{i} = \left\{\overline{\mathbf{u}}, \overline{\Phi}\right\}.
\end{equation}
To avoid large correlations in the data, we ensure that every sample is separated by almost one large eddy turnover time $t_{L} \sim L/U \sim (L^2/\epsilon)^{1/3}$, with $L$ the integral scale, from previous and following ones and construct a dataset from 80 equally spaced samples taken from 60000 iterations.
The learning dataset is then split into two parts so that the 60 first samples are dedicated to training and the remaining 20 samples are used to validate the model, which is useful especially to monitor the behavior of the learning phase. For the test dataset, we follow the same methodology and extract three different flow regimes used for the a-priori evaluation; a developed turbulence statistically similar to the training data, where both velocity and scalar source terms are active; a forced transitional regime to turbulence with fields initialized at large scales and a scalar decay driven by a forced turbulent flow. Reynolds number and scalar variance evolution as well as data separation from the datasets are shown in Figs. \ref{fig:train_data} and \ref{fig:tests_data} for training, validation and test respectively. 
In Fig. \ref{fig:dataset} we show 2{\sc d} slices at $z = \pi$ for different iterations $\mathcal{S}$ taken from the training dataset.

\subsection{Baseline models}
For benchmarking purposes, we consider two algebraic closure models \change{($\mathbf{s}_{\mathrm{DynSMAG}}, \mathbf{s}_{\mathrm{DynRG}}$)} and a recent model based on machine learning \change{($\mathbf{s}_{\mathrm{MLP}}$)}. Both algebraic models are based on a dynamic procedure using a test filter noted $\hat{\cdot}$ such that its filter length scale $\hat{\Delta} = 2 \bar{\Delta}$. The coefficient $C$ can then be determined using some information on the resolved scales with Lilly's method \cite{lilly1992proposed}.

The dynamic eddy diffusivity model \cite{moin1991dynamic} is an extension of the Smagorinsky model \cite{germano1991dynamic} for scalar transport. This is a functional model based on an eddy diffusivity leading to
\begin{equation}
    \mathbf{s}_{\mathrm{DynSMAG}} = C \bar{\Delta}^{2} ||\bar{S}|| \frac{\partial \overline{\Phi}}{\partial x_{j}},
\end{equation}
where $||\overline{S}||$ is the $L_{2}$ norm of the resolved strain rate tensor. This model is commonly used because it is stable when $C \in \mathbb{R}_{+}^{*}$, i.e., the  SGS scalar dissipation rate, $\mathbf{s}_{\mathrm{DynSMAG}} \cdot \mathbf{\nabla}\overline{\Phi}$, is always positive, meaning that the scalar-variance transfers are always from large to small scales.

The gradient model on the other side is a structural model based on a truncated series expansion using the filter scale \cite{leonard1974energy}. This model is known to provide a good local approximation and correlation with the exact residual flux as $\Delta$ tends to zero \cite{da2007analysis}. However, the model is also producing back-scatter, i.e., transfers from modeled to resolved scales which leads to instabilities due to the under-prediction of dissipation. To overcome this limitation, this model is often used in combination with an eddy diffusivity model \cite{clark1979evaluation}. Another recent improvement consists of a regularization of the model that eliminates back-scatter transfers \cite{balarac2013dynamic}. The dynamic regularized gradient model then writes
\begin{equation}
    \mathbf{s}_{\mathrm{DynRG}} =C \bar{\Delta}^{2} \bar{S}_{ij}^{\ominus} \frac{\partial \overline{\Phi}}{\partial x_{j}},
\end{equation}
where $\bar{S}_{ij}^{\ominus}$ only contains the direct energy transfer from the strain rate tensor.

We also consider a recently published NN SGS model \cite{portwood2021interpreting}, which relies on a Multilayer Perceptron (MLP) to predict the residual flux. This model uses the gradients of grid quantities, $\overline{\mathbf{u}}$ and $\overline{\Phi}$ as inputs and optimizes $L$ fully connected linear (or affine) layers with nonlinear ReLu activations $\mathcal{R}(x) = \max (0, x)$ such that 
\begin{equation}
    \mathbf{s}_{\mathrm{MLP}} = \mathbf{W}^{(L)}\mathbf{h}^{(L - 1)} + \mathbf{b}^{(L)}, \hspace{5mm} \mathbf{h}^{(i)} = \mathcal{R}\left(\mathbf{W}^{(i)}\mathbf{h}^{(i - 1)} + \mathbf{b}^{(i)}\right), \hspace{5mm} \mathbf{h}^{(0)} = \left\{\overline{\mathbf{u}}, \overline{\Phi}\right\},
\end{equation}
where $\mathbf{b}$ is a bias vector and $\mathbf{W}$ a weight matrix. To our knowledge, this model is the closest to our work.

\subsection{Metrics}
Most a priori evaluations in the literature characterize the performance of a model based on a structural metric which is defined by the error on the SGS term, and a functional metric that operates on the SGS residual flux. In particular, it is common to characterize the a priori performance of a given model by its ability to reproduce the distribution of SGS scalar dissipation, i.e., the transfer of energy between resolved and subgrid scales given by $\mathbf{s} \cdot \nabla \overline{\Phi}$. 
In this paper however, we focus on the modeling of the SGS term $\nabla \cdot \mathbf{s}$ and thus our proposed model does not have access to the residual flux $\mathbf{s}$. Therefore, we perform an extensive study of the predicted SGS term (or structural performance) statistical properties. As discussed in Ref. \cite{meneveau2000scale}, it is suggested that structural performance describes the short-term time evolution of a model and is of most importance to the understanding of local and instantaneous errors in a model. In this study, we evaluate three different types of metrics summarized in Table \ref{tbl:metrics} based on structural, physical and statistical considerations. The structural evaluation is given by the normalized root-mean-squared-error (RMSE) $\mathcal{L}_{\mathrm{rms}}$ and Pearson's cross-correlation coefficient $\mathcal{P}$ between the predicted SGS term and the filtered DNS. We also show the error on the integral of the SGS dissipation $\mathcal{I}$, which can be computed using the divergence of the SGS flux since
\begin{equation}
    \int_{V} \mathbf{s} \cdot \nabla \overline{\Phi} \,\, dV = - \int_{V} \overline{\Phi} \nabla \cdot \mathbf{s} \,\, dV
\end{equation}
if $V$ is our periodic domain.
Additionally, we consider statistical metrics to evaluate how consistent is the pdf of the predicted term w.r.t. the true one. The Jensen-Shannon distance $\mathcal{J}$ \cite{endres2003new} is a metric that measures the similarity between two probability distributions $P_{X}$, $P_{Y}$ defined on the same probability space $\mathcal{X}$ built on the Kullback-Leibler divergence $\mathcal{D}$, or relative entropy.
Also, we compute a Kolmogorov-Smirnov statistic test $\mathcal{K}$ that hypothesizes that two samples are drawn from the same distributions if its value 
is small ($\leq 0.1$) as shown in Ref. \cite{hodges1958significance}.

\begin{table*}[b]
\begin{ruledtabular}
\begin{tabular}{ll}
\textbf{Structural metric} & \\
\hline
Normalized RMSE & $\mathcal{L}_{\mathrm{rms}}(X, Y) = \sqrt{\frac{1}{N} ||X - Y||_{2}} / \sqrt{\langle Y^{2} \rangle - \langle Y \rangle^{2}}$\\
Pearson's coefficient & $\mathcal{P}(X, Y) = \mathrm{cov}(X, Y) / \sigma_{X} \sigma_{Y}$\\
\hline \hline
\textbf{Physical metric} & \\
\hline
Integral dissipation error & $\mathcal{I}(X, Y) = - \int_{V} \overline{\Phi} \nabla \cdot \mathbf{s}_{X} \,\, dV + \int_{V} \overline{\Phi} \nabla \cdot \mathbf{s}_{Y} \,\, dV$\\
\hline \hline
\textbf{Statistical metric} & \\
\hline
Kullback-Leibler divergence & $\mathcal{D}(P_{X} || P_{Y}) = \sum_i^N P_{X}(i) \ln \left( P_{X}(i) / P_{Y}(i) \right)$\\
Jensen-Shannon distance & $\mathcal{J}(P_{X} || P_{Y}) = \sqrt{\frac{1}{2} \mathcal{D}(P_{X} || M) + \frac{1}{2} \mathcal{D}(P_{Y} || M)}, M = \frac{1}{2}(P_{X} + P_{Y})$\\
Kolmogorov-Smirnov statistic & $\mathcal{K}(X, Y) = \sup_{t} | F_{X}(t) - F_{Y}(t) |, F_{A}(t) = \frac{1}{N} \sum_i^N \mathbf{1}_{A_i \leq t}$
\end{tabular}
\end{ruledtabular}
\caption{Equations for the structural, physical and statistical metrics used to evaluate the a priori performance of the different models. \label{tbl:metrics}}
\end{table*}

\section{Transformation-invariant learning framework}
In this section, we describe how we model the SGS term, $\nabla \cdot \mathbf{s}$, using a NN model, referred to as $\mathcal{M}_{\mathrm{NN}}$, where the objective is to approximate
\begin{equation}
    \mathcal{M}_{\mathrm{NN}}(\overline{\mathbf{u}}, \overline{\Phi}) \approx \nabla \cdot \mathbf{s}.
\end{equation}
We choose four nonexhaustive physical relationships that must be verified by any SGS scalar flux model and show how to include them within a NN as hard and soft constraints. The first three constraints are based on frame symmetry \cite{berselli2006mathematics} which encompass Galilean, translation and rotation invariance. The last constraint ensures the linearity of the transport equation, required in order to generalizes to different scalar quantities. 

\subsection{Translation invariance}
Translation invariance states that the considered SGS NN model shall not be location-dependent. In other words,  model $\mathcal{M}_{\mathrm{NN}}$ shall satisfy the following constraint
\begin{equation}
    \forall \delta \in \mathbb{R}, \hspace{5mm} \mathcal{M}_{\mathrm{NN}}(T_{\delta} \overline{\mathbf{u}}, T_{\delta} \overline{\Phi}) = T_{\delta} \mathcal{M}_{\mathrm{NN}}(\overline{\mathbf{u}}, \overline{\Phi}),
    \label{eq:translation}
\end{equation}
where $T_\delta$ is a translation operator for spatial displacement $\delta$ such that $T_\delta \mathbf{y}(\mathbf{x}) = \mathbf{y}(\mathbf{x} + \delta)$. In the deep learning literature, in contrast to fully connected architectures which refers to NN architectures based on dense or fully connected layers, e.g., the MLP used in Ref. \cite{portwood2021interpreting}, Convolutional Neural Networks (CNNs) are described by a combination of convolution and activation layers. Since the convolution commutes with translation, this type of architecture provides a built-in invariance to spatial displacements, which satisfy \eqref{eq:translation}. We note that using a different set of inputs can already ensure some symmetries of the residual flux \cite{speziale1985galilean} but reduces the flexibility of the model. 

Convolutional architectures also appear as a natural choice when dealing with structured tensors (i.e., time, space, space-time processes) as they relate to local filtering operations with respect to tensor dimensions. For instance, convolutional layers embed all discretized-filtering operations, such as first-order or higher-order derivatives of the considered variables. This may provide the basis for some physical interpretation of CNN architectures. 
Importantly, CNNs greatly outperform fully connected architectures and are currently the state-of-the-art schemes for a wide range of machine learning applications with one-dimensional \cite{ince2016real}, two-dimensional \cite{krizhevsky2012imagenet}, or three-dimensional \cite{kamnitsas2017efficient} states. 

Let us call now $\mathcal{M}_{\mathrm{CNN}}$ a model built from convolutions and nonlinear activation layers as presented above. This model is expected to perform well on the metric $\mathcal{L}$ for which it has been optimized but should not strictly or even numerically (as shown in Table \ref{tbl:invariances} for some invariances except the translational one) verify well-established physical symmetries and properties. From now on, we will refer to the transformation-invariant NN model as $\mathcal{M}_{\mathrm{SGTNN}}$ for "Subgrid Transport Neural Network."

\subsection{Transport linearity}
\begin{figure*}[t]
  \centering
  \includegraphics[width=.5\textwidth]{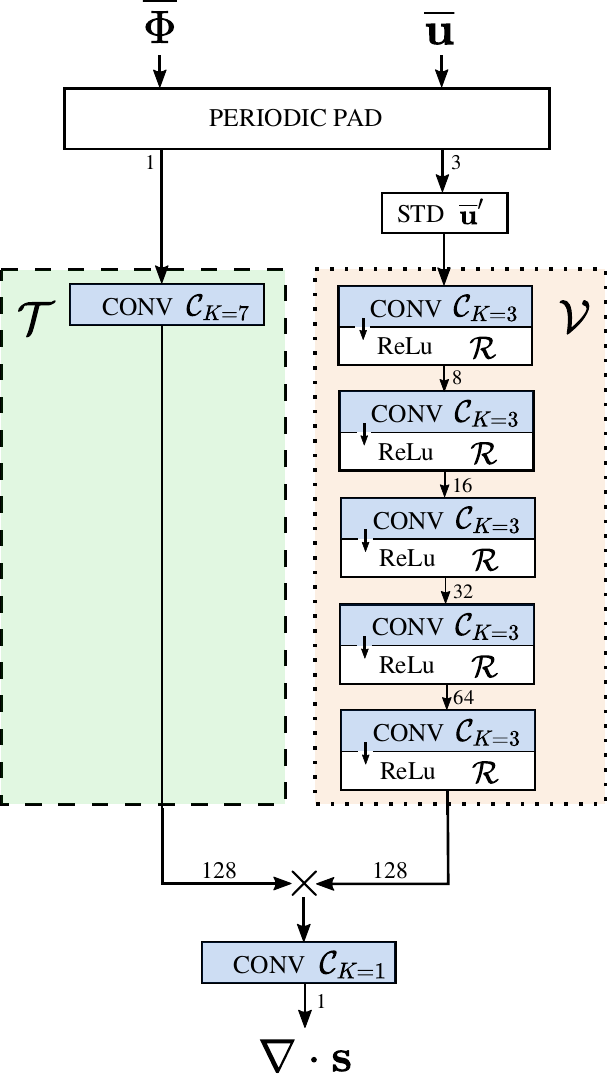}
  \caption{Illustration of the subgrid transport neural network architecture (SGTNN). At the top, two nonlearnable steps are applied; the periodic pad replicates boundaries according to the kernel width of convolution kernel and std transforms the velocities to ensures Galilean invariance. Then, the two submodels $\mathcal{V}$ (right) for the velocity and $\mathcal{T}$ (left) for the scalar transport are represented in dotted and dashed lines, respectively. The convolution with kernel size $K = 1$ corresponding to $\mathcal{C}^{+}$ applied to the combined result of the latter is depicted at the bottom. The number of filters within the NN is shown along with the arrows.
  \label{fig:SGTNN}
  }
\end{figure*}

The linearity of the scalar transport equation is a fundamental property that shall be verified by the proposed SGS model. 
Formally, it comes to verify the following constraint
\begin{equation}
    \forall \lambda \in \mathbb{R}, \hspace{5mm} \mathcal{M}_{\mathrm{NN}}(\overline{\mathbf{u}}, \lambda \overline{\Phi}) = \lambda \mathcal{M}_{\mathrm{NN}}(\overline{\mathbf{u}}, \overline{\Phi}).
    \label{eq:linearity}
\end{equation}
This relationship can not be verified within $\mathcal{M}_{\mathrm{CNN}}$ since the model contains nonlinearities introduced by the $\mathcal{R}$ operator. We propose to split the model $\mathcal{M}_{\mathrm{CNN}}$ into two submodels $\mathcal{V}$ and $\mathcal{T}$ that take as inputs the velocities and the transported scalar, respectively. $\mathcal{V}$ involves a classic CNN architecture. By contrast, to conform to the linearity property, model $\mathcal{T}$ applies a single convolution layer with no bias and nonlinear activation. 
The resulting SGS model is given by
\begin{equation}
    \mathcal{M}_{\mathrm{SGTNN}}(\overline{\mathbf{u}}, \overline{\Phi}) = \left[ \mathcal{V}(\overline{\mathbf{u}}) \times \mathcal{T}(\overline{\Phi}) \right] * \mathcal{C}^{+},
\end{equation}
where $*$ is the discrete convolution operator, $\times$ is a term-by-term matrix product, and $\mathcal{C}^{+}$ is the convolution kernel with unitary width $K=1$. We sketch the resulting architecture in Fig. \ref{fig:SGTNN}. 

First, the periodic pad replicates boundaries according to the kernel width of convolutions kernels $\mathcal{C}$. The second stage applies in parallel operators $\mathcal{V}$ and $\mathcal{T}$ and computes the term-by-term product of their outputs, which are $128 \times 128 \times 128$ tensors. The last convolutional layer maps the resulting 128-dimensional representation to a scalar field using a kernel width of unitary size. 
We can now show that the relationship \eqref{eq:linearity} holds:
\begin{align*}
    \mathcal{M}_{\mathrm{SGTNN}} \left( \overline{\mathbf{u}}, \lambda \overline{\Phi} \right)
    = \sum_{i = 0}^{d} \mathcal{V}(\overline{\mathbf{u}})_{i} \lambda \mathcal{T}(\overline{\Phi})_{i} C_{i}^{+} 
    = \lambda \sum_{i = 0}^{d} \mathcal{V}(\overline{\mathbf{u}})_{i} \mathcal{T}(\overline{\Phi})_{i} C_{i}^{+},
\end{align*}
where $d$ is the input dimension of $\mathcal{C}^{+}$ (here $d = 128$) and linearity of $\mathcal{T}$ is implied from its convolution. Finally,
\begin{align*}
    \mathcal{M}_{\mathrm{SGTNN}} \left( \overline{\mathbf{u}}, \lambda \overline{\Phi} \right)
    &= \lambda  \left( \left[ \mathcal{V}(\overline{\mathbf{u}}) \times \mathcal{T}(\overline{\Phi}) \right] * \mathcal{C}^{+} \right)\\
    &= \lambda \mathcal{M}_{\mathrm{SGTNN}}(\overline{\mathbf{u}}, \overline{\Phi}),
\end{align*}
which validates that the architecture $\mathcal{M}_{\mathrm{SGTNN}}$ has a linear path on $\overline{\Phi}$.

\subsection{Galilean invariance}
The next constraint is based on frame invariance. It is known that the Navier-Stokes and the transport equations as well as their filtered form are Galilean-invariant \cite{speziale1985galilean}. In order to fulfill this property on the modeled scales, we must ensure that our SGS model is also form-invariant under the Galilean group of transformations. In particular, we want to have a description of turbulence that is the same in all inertial frames of reference. This translates easily to our problem as
\begin{equation}
    \forall \beta \in \mathbb{R}, \hspace{5mm} \mathcal{M}_{\mathrm{NN}}(\overline{\mathbf{u}} + \beta, \overline{\Phi}) = \mathcal{M}_{\mathrm{NN}}(\overline{\mathrm{u}}, \overline{\Phi}).
\end{equation}
We propose to solve this problem by reducing the velocity to a standardized quantity, i.e, with zero mean $\langle (\overline{\mathbf{u}} + \beta)^{\prime} \rangle = 0$ for each instantaneous sample. Let us denote the standardized operator $\cdot^{\prime}$, and we have
\begin{align*}
    \left( \overline{\mathbf{u}} + \beta \right)^{\prime} &= \overline{\mathbf{u}} + \beta - \langle \overline{\mathbf{u}} + \beta \rangle \\
    &= \overline{\mathbf{u}} - \langle \overline{\mathbf{u}} \rangle \\
    &= \left( \overline{\mathbf{u}} \right)^{\prime}.
\end{align*}
Note that this assumption is true only if bias is removed from the operator $\mathcal{C}$. We include this standardization step after the periodic pad and apply it on the velocities prior to the submodel $\mathcal{V}$.

\subsection{Rotation invariance}
The last fundamental invariance required to fulfill frame symmetry is related to reflections and rotations \cite{oberlack1997invariant}. The filtered equation \eqref{eq:filtered_advection} holds under the rotation of the coordinate system and velocity vector, 
\begin{equation}
    \mathcal{M}_{\mathrm{NN}}(A_{ij} \overline{\mathbf{u}}_{j}, \overline{\Phi}) = \mathcal{M}_{\mathrm{NN}}(\overline{\mathbf{u}}, \overline{\Phi}),
    \label{eq:rotation_invariance}
\end{equation}
for any rotation matrix $A$ with $A^{T}A = AA^{T} = I$ in a rotated frame $\mathbf{y}_{i} = A_{ij}\mathbf{y}_{j}$. Ensuring rotation invariance in a NN is a difficult problem since convolution kernels are not imposed but rather learnt by the optimization algorithm. Some progress has been made to exploit these symmetries \cite{cohen2016group, weiler20183d} but are either limited to finite subsets of rotations or require major changes and priors in the architecture. A common and flexible way to approximate this invariance is to use data augmentation, i.e., extending the dataset with new samples that exhibit the desired invariance. We used permutations of our initial data that verifies \eqref{eq:rotation_invariance} with angles of 90$^{\circ}$.

\subsection{NN architecture and learning scheme}
The optimization problem is formulated as the minimization of an energy term $\mathcal{L}$ also called a loss function which in our case will be defined as a simple mean-squared error ($L_{2}$) on the SGS term prediction. Obtaining the gradients of $\mathcal{L}$ is done through automatic differentiation, which is available directly in most NN frameworks. Finally, the NN models optimize the parameter space w.r.t
\begin{equation}
    \mathcal{L}\left[ \mathcal{M}_{\mathrm{NN}}(\theta_{k}), \nabla \cdot \mathbf{s} \right] = \left\lVert \mathcal{M}_{\mathrm{NN}}(\theta_{k}) - \nabla \cdot \mathbf{s} \right\rVert_{2},
\end{equation}
where $\theta_{k}$ is the model input. Since our simulations are done in a periodic domain, we can replicate periodically our input at the boundaries without introducing any error. This allows us to remove any padding inside the NN.

After extensive experimentation, the best results in our scenarios were obtained using a velocity submodel $\mathcal{V}$ composed of six nonlinear convolution units, i.e., applying $\mathcal{R}$ after $\mathcal{C}$ with kernels of size $3 \times 3 \times 3$ and increasing the number of filters from $8$ to $128$. The transport submodel $\mathcal{T}$ is built from a single convolution with a wider kernel of size $7 \times 7 \times 7$ cleared from any nonlinearities.
The minimization problem is carried using an Adam optimizer \cite{kingma2015adam} for 1000 epochs. We use an adaptive step-based scheduler, which decreases the learning-rate from $1e^{-4}$ by a factor $\rho = 0.75$ every 250 epochs. As a pre-processing step, we normalize the input data to zero mean and unitary variance, which has been shown to improve the convergence speed of gradient descent steps \cite{ioffe2015batch}.

The implementation using PyTorch is available at Ref. \cite{frezat2020sgtnn}, and a priori results presented below can be reproduced with the given notebooks.

\section{Results}
We first show in Table \ref{tbl:invariances} numerical evidence of the unrealistic a priori predictions from the $\mathcal{M}_{\mathrm{CNN}}$ w.r.t the considered physical invariances. It is demonstrated that in practice, the $\mathcal{M}_{\mathrm{SGTNN}}$ strictly enforces scalar linearity and Galilean invariance with zero variance over multiple runs of uniformly sampled $\lambda$ and $\beta$ coefficients. The soft constraint imposed for rotation invariance reduces the error variance from one order of magnitude over the six tested permutations.

\begin{table}[t]
\begin{ruledtabular}
\begin{tabular}{lcccc}
 & \multicolumn{2}{c}{$\mathcal{M}_{\mathrm{CNN}}$} & \multicolumn{2}{c}{$\mathcal{M}_{\mathrm{SGTNN}}$} \\
 & E[$\mathcal{L}_{\mathrm{rms}}$] & Var[$\mathcal{L}_{\mathrm{rms}}$] & E[$\mathcal{L}_{\mathrm{rms}}$] & Var[$\mathcal{L}_{\mathrm{rms}}$] \\
\hline
$\mathcal{M}_{\mathrm{NN}}(\overline{\mathbf{u}}, \lambda \overline{\Phi})$ & 1.0238 & 0.0945 & 0.8356 & 0.0\\
$\mathcal{M}_{\mathrm{NN}}(\overline{\mathbf{u}} + \beta, \overline{\Phi})$ & 0.9330 & 0.0022 & 0.8356 & 0.0\\
$\mathcal{M}_{\mathrm{NN}}(A_{ij} \overline{\mathbf{u}}, \overline{\Phi})$ & 0.8741 & $1.4540 \times 10^{-7}$ & 0.8355 & $2.044 \times 10^{-8}$\\
\end{tabular}
\end{ruledtabular}
\caption{Evaluation of the three additional constraints provided by $\mathcal{M}_{\mathrm{SGTNN}}$; expectation and variance of the normalized root-mean-squared error on the SGS term predicted by NN models from many realizations on the testing data. Each row checks the transport linearity, Galilean, and rotation invariance, respectively. Note that $\lambda$ and $\beta$ are uniformly sampled values. \label{tbl:invariances}}
\end{table}

The presented models are now evaluated on different scenarios in two ways: first, we evaluate the performance of a model a priori using the metrics described in the previous section, then we run simulations with the surrogate models and evaluate their evolution through time, i.e., a posteriori.

\subsection{A priori evaluation}

\subsubsection{Developed turbulence regime}

\begin{table*}[t]
\begin{ruledtabular}
\begin{tabular}{lccccc}
& $\downarrow \mathcal{L}_{\mathrm{rms}}(X, Y)$ & $\uparrow \mathcal{P}(X, Y)$ & $\downarrow \mathcal{J}(P_{X} || P_{Y})$ & $\downarrow \mathcal{K}(X, Y)$ & $\downarrow \mathcal{I}(X, Y)$\\
\hline
$\mathcal{M}_{\mathrm{DynSmag}}$ & 0.9400 & 0.3602 & 0.2840 & 0.1742 & 0.1767\\
$\mathcal{M}_{\mathrm{DynRG}}$ & 0.8748 & 0.4969 & 0.3471 & 0.2137 & \textbf{0.0624}\\
$\mathcal{M}_{\mathrm{MLP}}$ & 1.2512 & 0.1365 & \textbf{0.0449} & \textbf{0.0159} & 0.0896\\
$\mathcal{M}_{\mathrm{CNN}}$ & 0.8734 & 0.5597 & 0.0868 & 0.0554  & 0.1021\\
$\mathcal{M}_{\mathrm{SGTNN}}$ & \textbf{0.8356} & \textbf{0.6118} & 0.1070 & 0.0690 & 0.1190
\end{tabular}
\end{ruledtabular}
\caption{A priori evaluation of the SGS term in developed turbulence regime using testing data with the metrics described in Section V.1. \label{tbl:apriori_dev}}
\end{table*}

\begin{figure*}[t]
  \centering
  \begin{minipage}[c]{.32\textwidth}
    \centering
    \includegraphics[width=1.\linewidth]{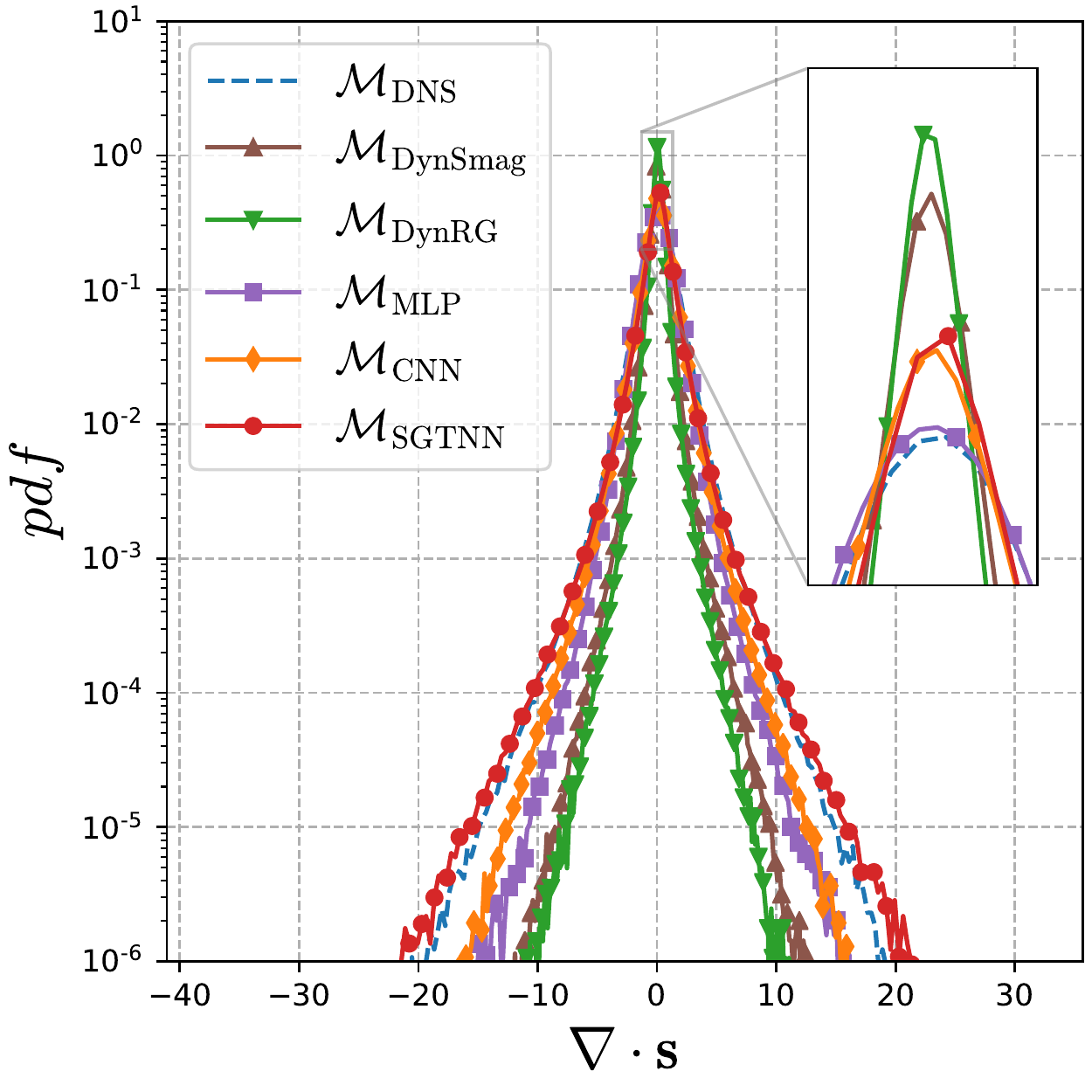}
  \end{minipage}
  \begin{minipage}[c]{.32\textwidth}
    \centering
    \includegraphics[width=1.\linewidth]{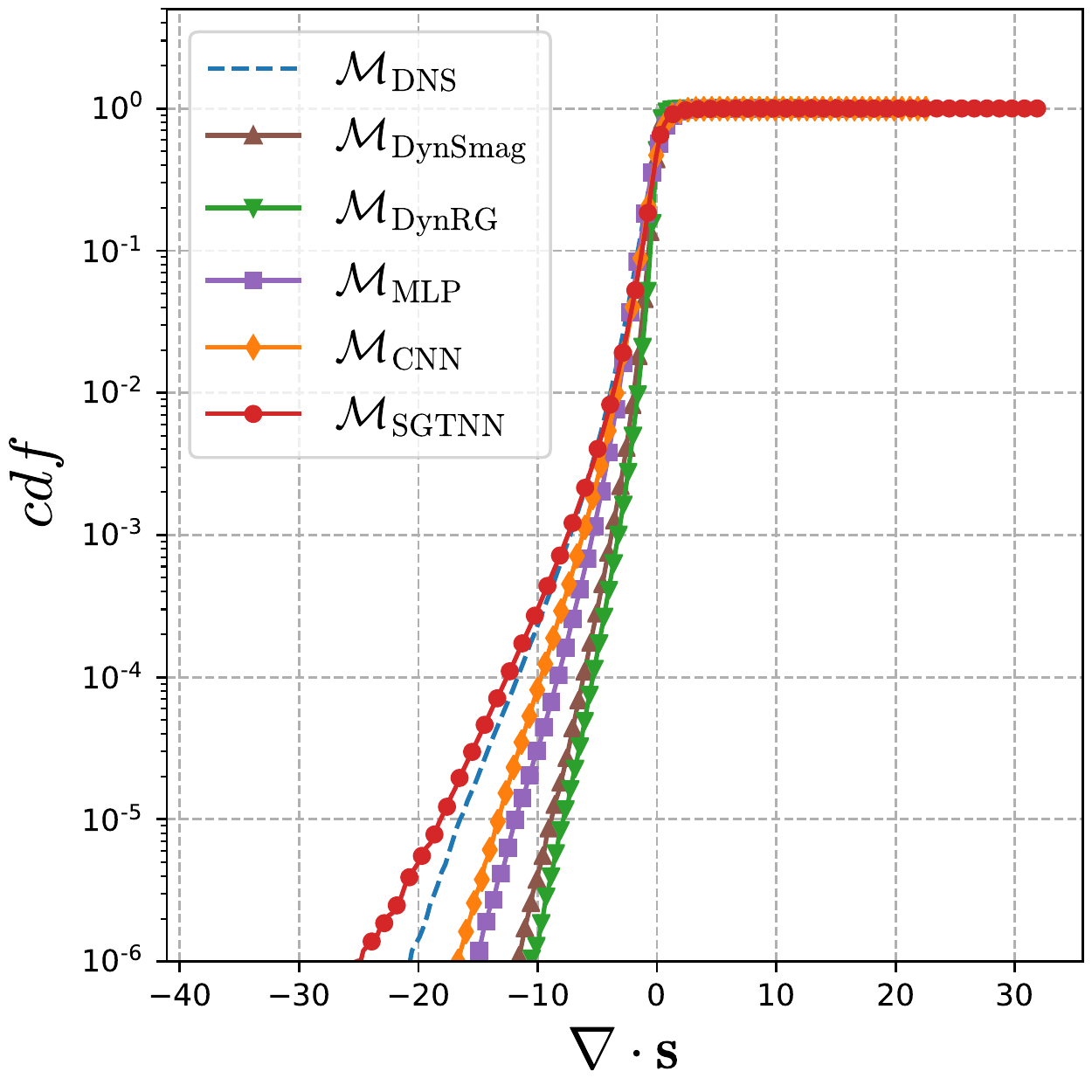}
  \end{minipage}
  \begin{minipage}[c]{.32\textwidth}
    \centering
    \includegraphics[width=1.\linewidth]{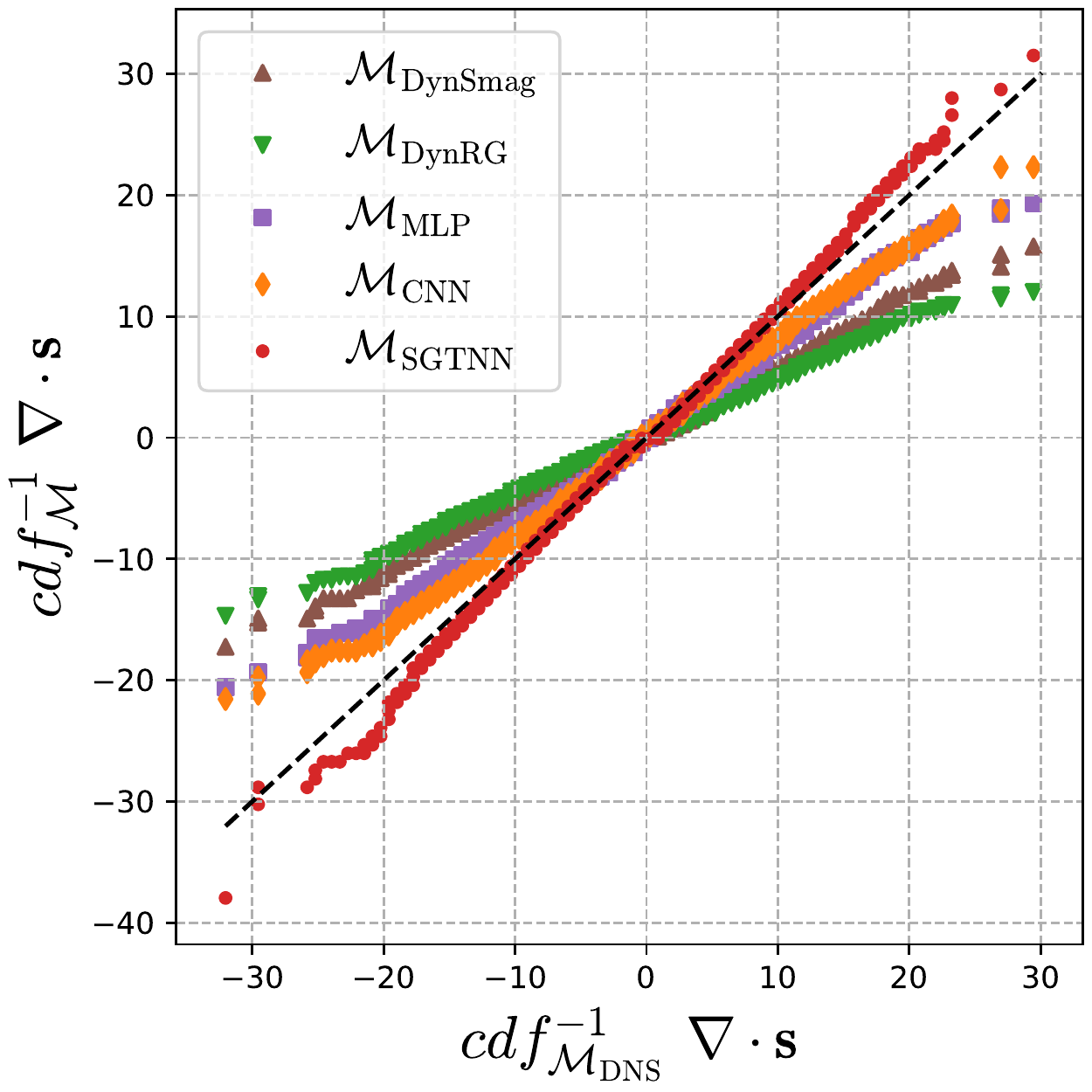}
  \end{minipage}
  \caption{Probability distribution function (left), cumulative distribution function (middle), and quantiles-quantiles (right) of the SGS term in developed turbulence regime over the entire testing data. Both distribution function tails are made explicitly visible using log scale on the y-axis.
  \label{fig:apriori_dev_stats}
  }
\end{figure*}

We first perform the evaluation on the developed turbulence regime of the testing dataset (see Fig. \ref{fig:tests_data}) which is the closest to the training data, \change{and where NN models are expected to perform the best}. 
\change{It is important to note that some metrics such as the mean-squared-error and correlation coefficient strongly depends on the kernel used to filter out the small scales features, as shown in Ref. \cite{fabre2011development}.}
In Table \ref{tbl:apriori_dev} we show the results of the a priori evaluation with the same range of values as the training data, i.e. $\lambda = 1$ and $\beta = 0$. The invariant model $\mathcal{M}_{\mathrm{SGTNN}}$ gives the most consistent structural results compared to the DNS. In particular, we note a substantial improvement on the $\mathcal{L}_{\mathrm{rms}}$ and $\mathcal{P}$ compared to the $\mathcal{M}_{\mathrm{CNN}}$. On the statistical metrics, it is clear from the inset in Fig. \ref{fig:apriori_dev_stats} that the $\mathcal{M}_{\mathrm{MLP}}$ is better at reproducing mean values (around $\nabla \cdot \mathbf{s} = 0$), which tends to lead to a smaller $\mathcal{J}$ and $\mathcal{K}$. However, the $\mathcal{M}_{\mathrm{SGTNN}}$ is more accurate on the tails of the distribution compared to the other models. This is particularly emphasized by the Quantile-Quantile (QQ) plot, which draws the theoretical quantiles (DNS) on the x-axis and the predicted ones on the y-axis (Fig. \ref{fig:apriori_dev_stats}, right). Again, it is found that the $\mathcal{M}_{\mathrm{SGTNN}}$ is in best agreement with the theoretical quantiles, for which a perfectly reproduced distribution would fit a straight line. 
\change{During experimentation, we also noticed that even if the best results of $\mathcal{M}_{\mathrm{SGTNN}}$ are achieved with every discussed invariance, the scalar linearity shows the most noticeable impact on a priori performance.}
Since this regime is similar to the training data, models based on machine learning are expected to give optimal results in the term of the target metric $\mathcal{L}$. We see, however, that the $\mathcal{M}_{\mathrm{MLP}}$ is not performing well when evaluated on the divergence. This could be explained by the fact that it is minimizing the error on the residual flux rather than the SGS term itself. 
The next two \change{regimes} will test the ability of the NN-based models to generalize to flows that were not part of the training data.

\subsubsection{Transitional regime}

\begin{figure*}[t]
  \centering
  \begin{minipage}[c]{.49\textwidth}
    \centering
    \includegraphics[width=1.\linewidth]{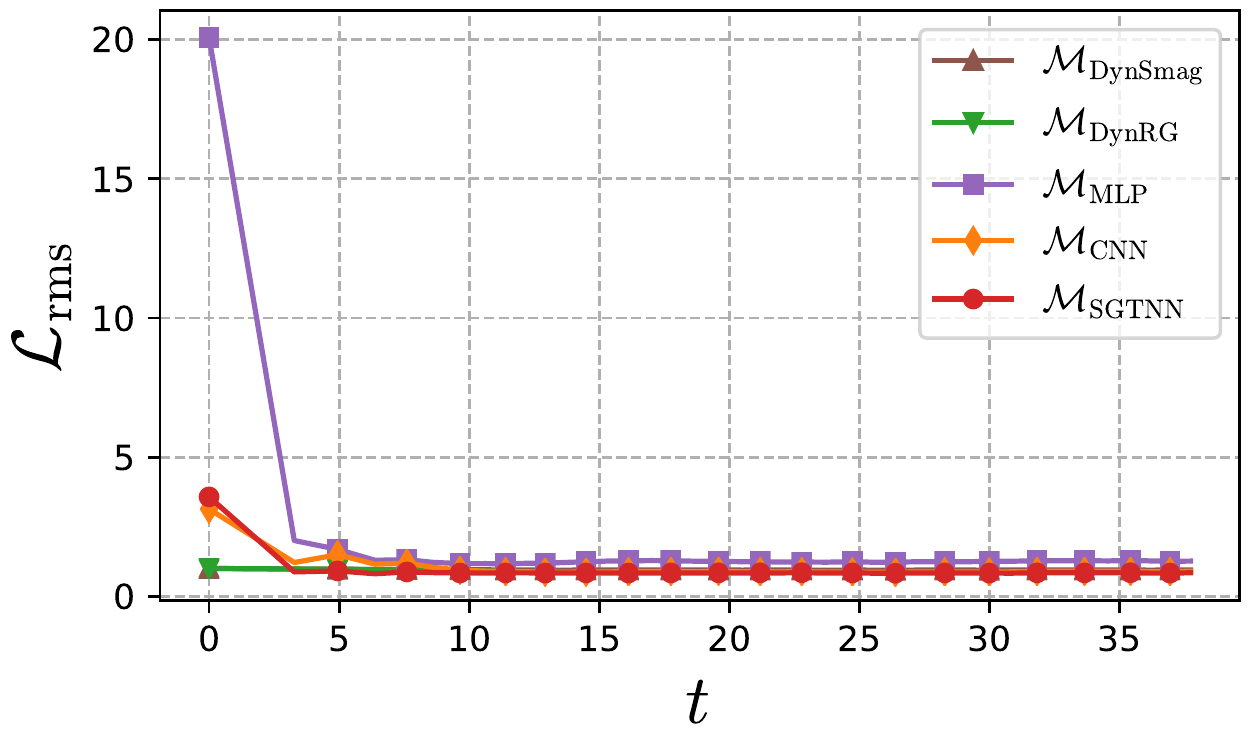}
  \end{minipage}
  \begin{minipage}[c]{.49\textwidth}
    \centering
    \includegraphics[width=1.\linewidth]{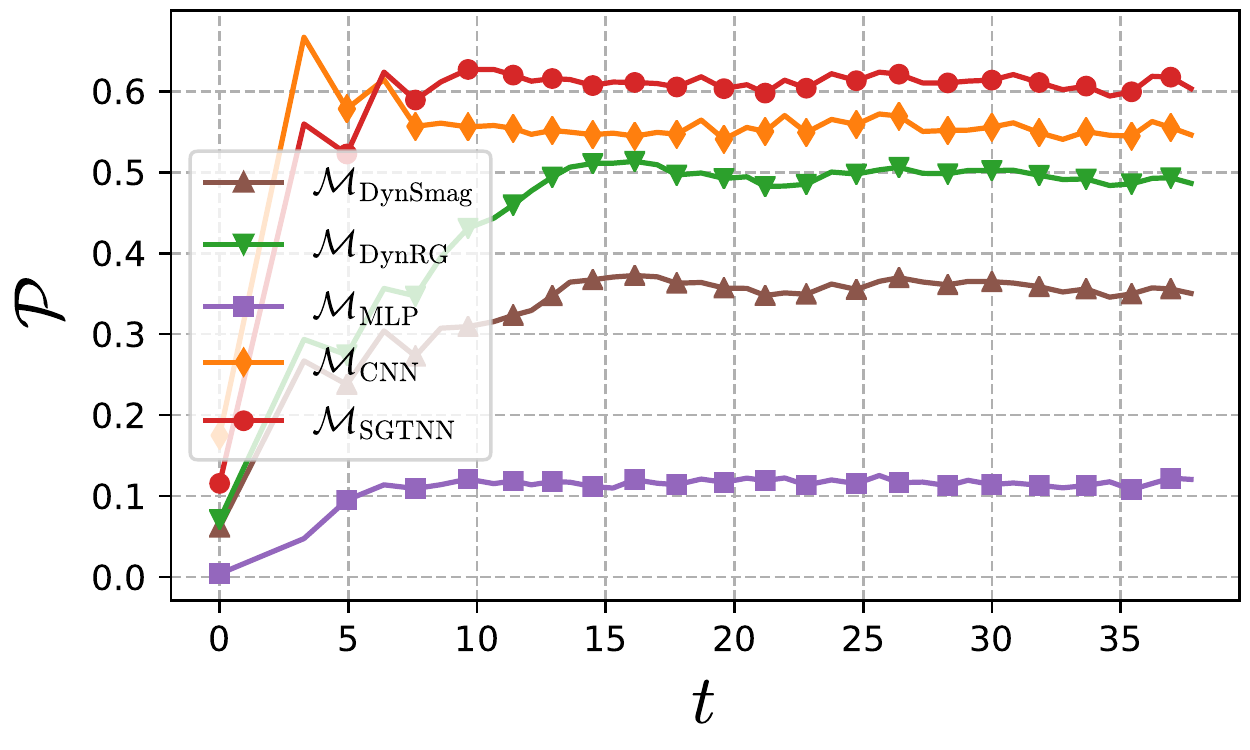}
  \end{minipage}
  \begin{minipage}[c]{.49\textwidth}
    \centering
    \includegraphics[width=1.\linewidth]{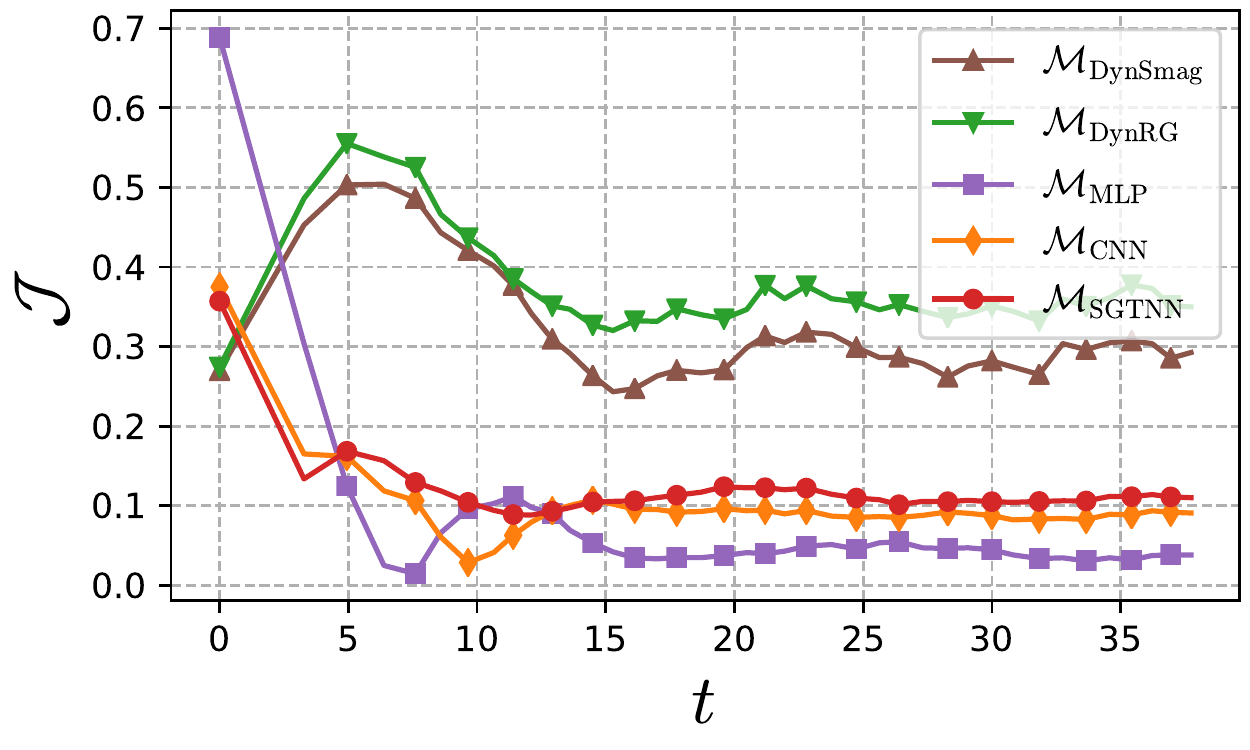}
  \end{minipage}
  \begin{minipage}[c]{.49\textwidth}
    \centering
    \includegraphics[width=1.\linewidth]{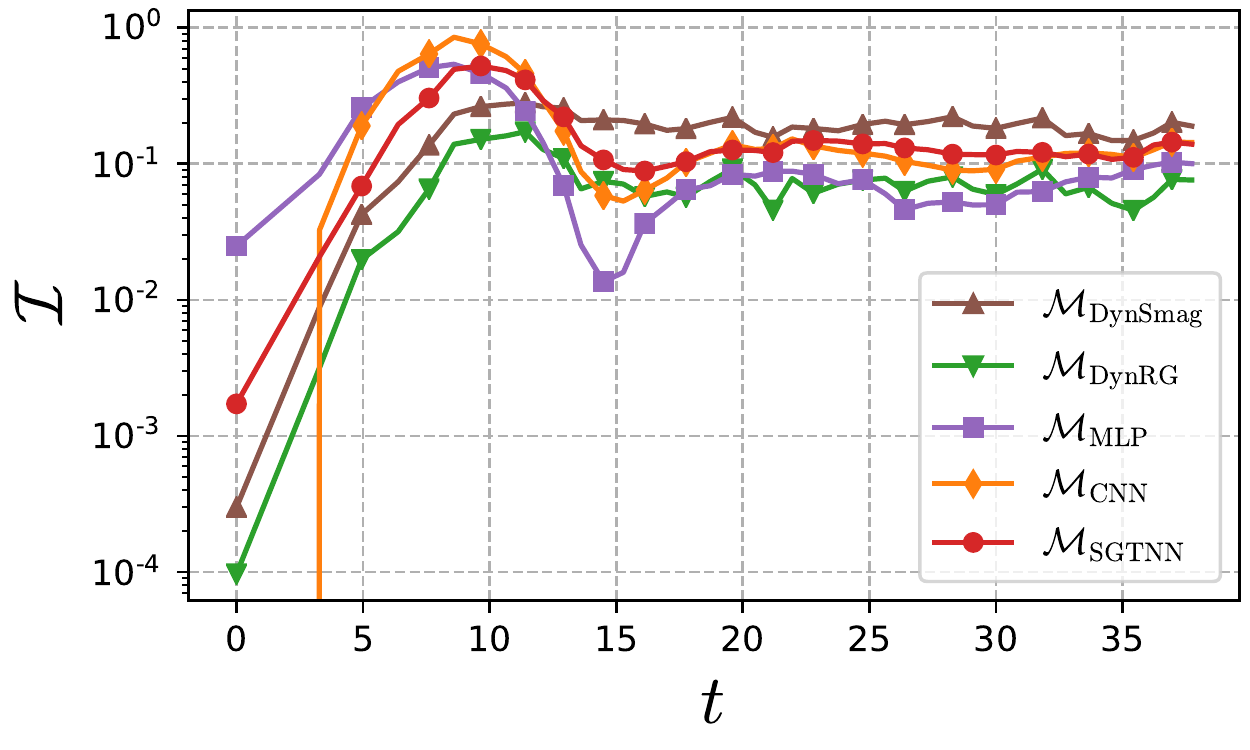}
  \end{minipage}
  \caption{Time evolution of the different metrics in transitional regime. Normalized root-mean-squared error (top left), Pearson's coefficient (top right), Jensen-Shannon distance (bottom left), and integral dissipation error (bottom right).
  \label{fig:apriori_transitional_stats}
  }
\end{figure*}

In this regime, we look at a forced flow that transitions from newly initialized to turbulent motion, as seen in Fig. \ref{fig:tests_data}. In the a priori tests, we consider both velocities and scalar in the transitional regime. Fig. \ref{fig:apriori_transitional_stats} shows that the $\mathcal{L}_{\mathrm{rms}}$ is large for the first sample prediction of NN models but rapidly decreases to a similar behavior as in developed turbulence. The correlation given by $\mathcal{P}$ is also quite low during the first iterations but quickly reaches its statistical mean at $t \approx 5$, as well as the metric $\mathcal{J}$ from which a large distance is observed especially from the $\mathcal{M}_{\mathrm{MLP}}$. 
\begin{figure*}[t]
  \centering
  \includegraphics[width=1.0\textwidth]{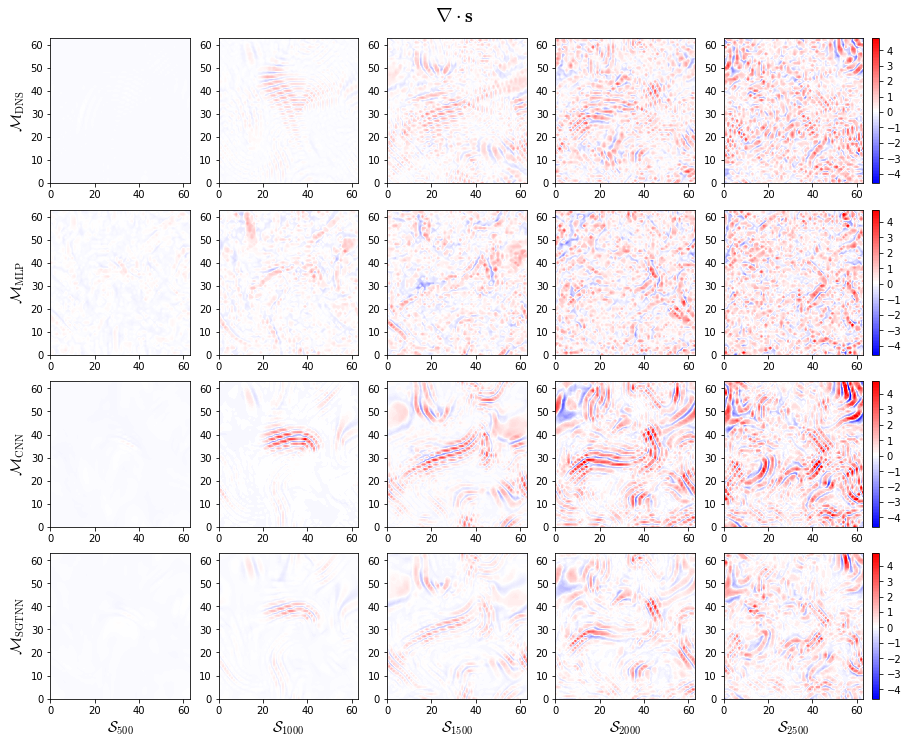}
  \caption{Data slices of the predicted SGS terms from NN models in transitional regime. It is particularly visible that $\mathcal{M}_{\mathrm{MLP}}$ is producing a non-null SGS term during the first iterations, while $\mathcal{M}_{\mathrm{CNN}}$ quickly produces large localized values.
  \label{fig:slices_transitional}
  }
\end{figure*}
We can also see in Fig. \ref{fig:slices_transitional} that the first iterations of the $\mathcal{M}_{\mathrm{MLP}}$ and $\mathcal{M}_{\mathrm{CNN}}$ are producing a non-null SGS term while it is expected to be almost zero, since the presence of turbulence is limited at that time. On the other hand, the $\mathcal{M}_{\mathrm{SGTNN}}$ follows an expected behavior, since it gradually produces a SGS term that corresponds to the DNS. A difference that appears for each NN-based model is that they produce a higher dissipation error spike compared the other models. This can be explained by the fact that none of these models have been optimized nor constrained on the scalar variance transfer. Ref. \cite{vollant2017subgrid} also proposes to build a model based on multi-objective minimization, which aims to optimize both functional and structural performances and could also be applied to our model.

\subsubsection{Scalar decay regime}

\begin{figure*}[t]
  \centering
  \begin{minipage}[c]{.49\textwidth}
    \centering
    \includegraphics[width=1.\linewidth]{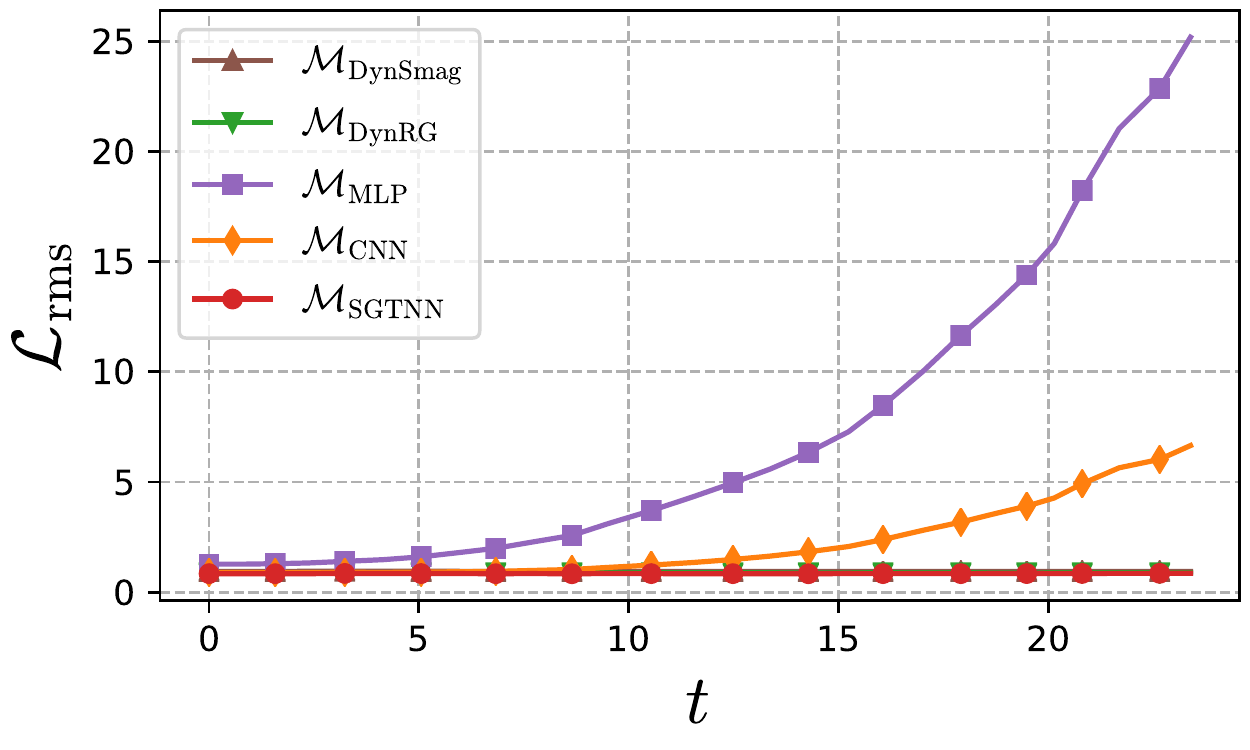}
  \end{minipage}
  \begin{minipage}[c]{.49\textwidth}
    \centering
    \includegraphics[width=1.\linewidth]{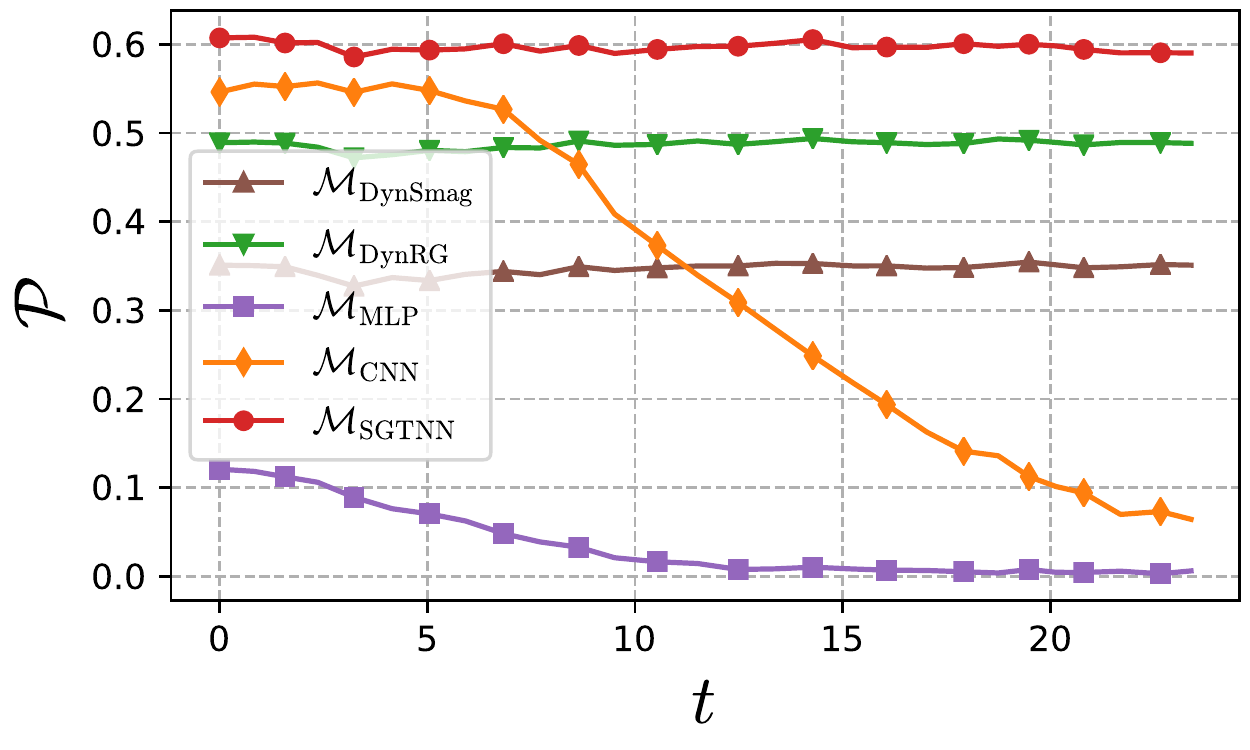}
  \end{minipage}
  \begin{minipage}[c]{.49\textwidth}
    \centering
    \includegraphics[width=1.\linewidth]{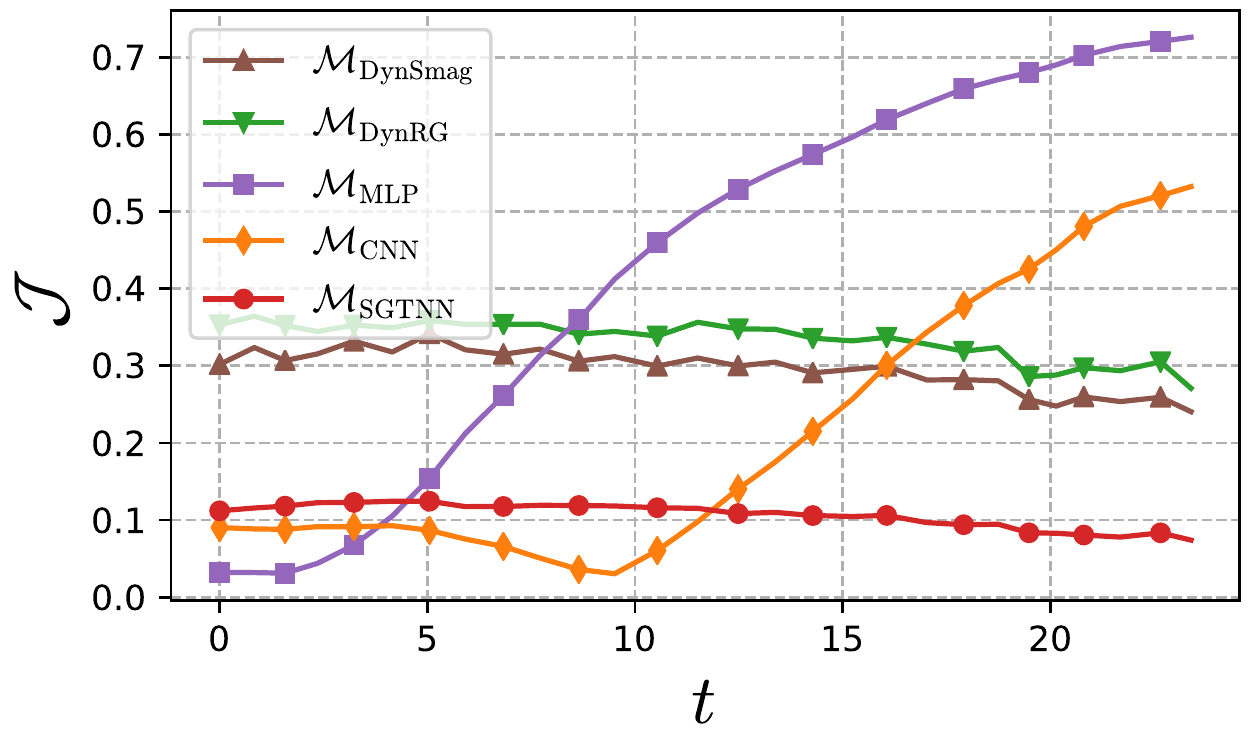}
  \end{minipage}
  \begin{minipage}[c]{.49\textwidth}
    \centering
    \includegraphics[width=1.\linewidth]{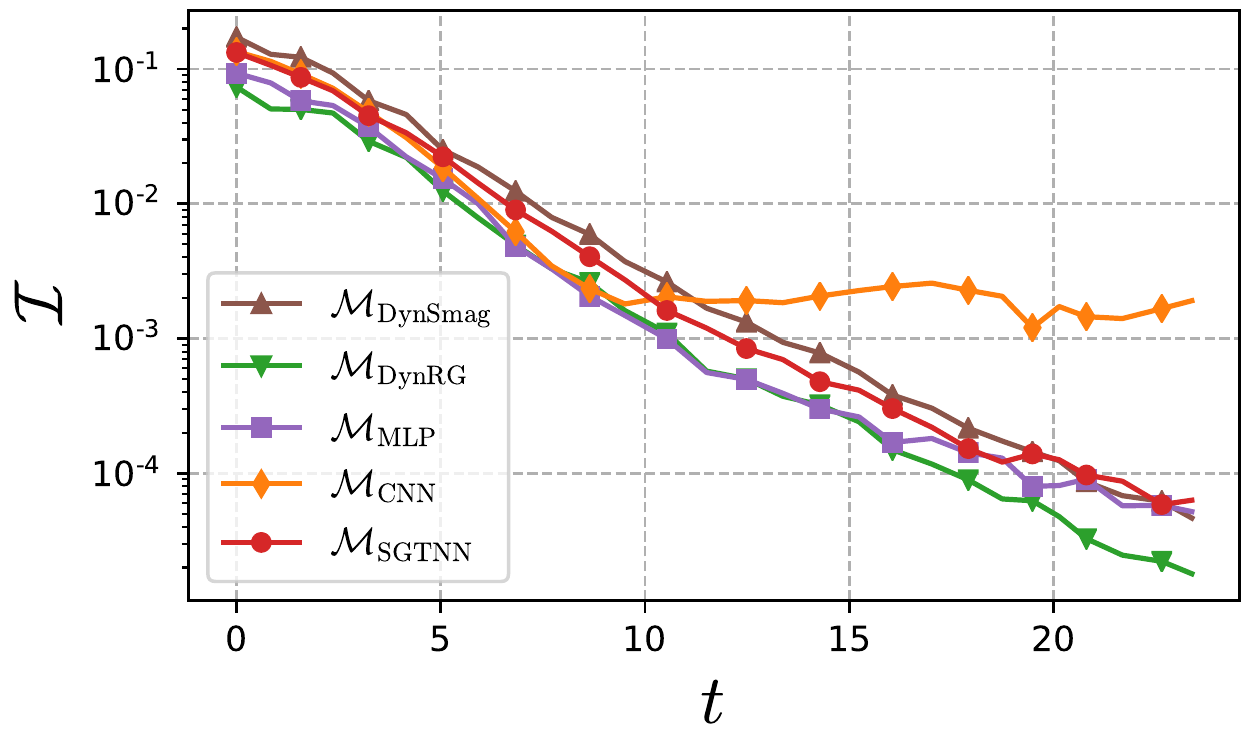}
  \end{minipage}
  \caption{Time evolution of the different metrics in scalar decay regime. Normalized root-mean-squared error (top left), Pearson's coefficient (top right), Jensen-Shannon distance (bottom left), and integral dissipation error (bottom right).
  \label{fig:apriori_scalar_decay_stats}
  }
\end{figure*}

\begin{figure*}[t]
  \centering
  \includegraphics[width=1.0\textwidth]{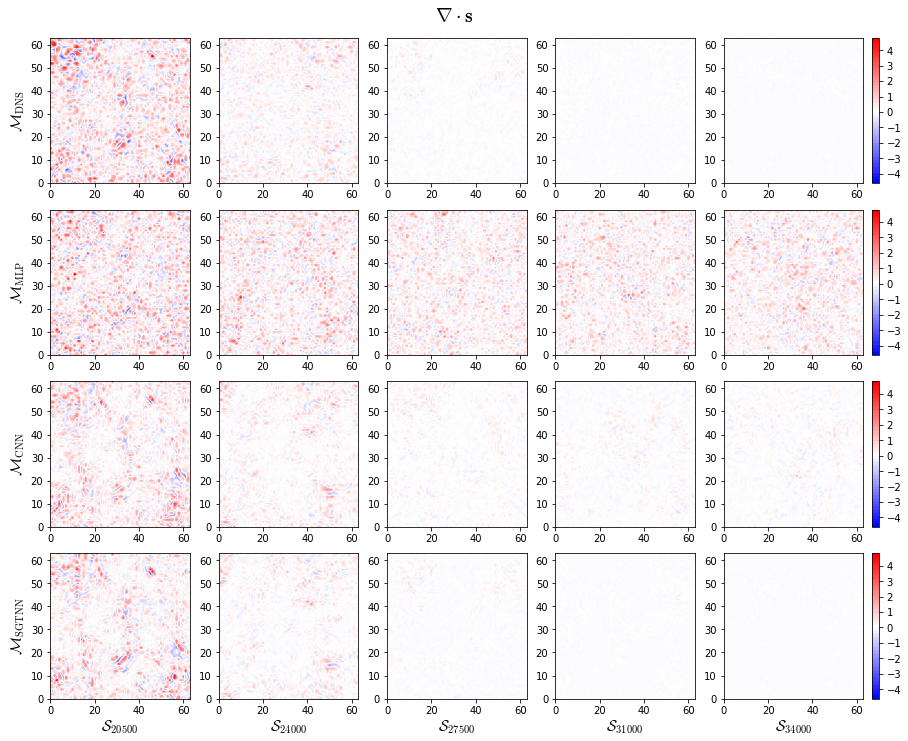}
  \caption{Data slices of the predicted SGS terms from NN models in scalar decay regime. It is particularly visible that $\mathcal{M}_{\mathrm{MLP}}$ and $\mathcal{M}_{\mathrm{CNN}}$ are still predicting a non-null SGS term when the scalar is completely mixed.
  \label{fig:slices_scalar_decay}
  }
\end{figure*}

In this last a priori \change{regime}, we remove the scalar forcing while maintaining the velocity field in turbulent motion. This results in a decay of the scalar within the domain. This regime is a difficult case for the models presented in this work and assesses their capacity to generalize to flow conditions that have not been seen during the training phase. Both $\mathcal{M}_{\mathrm{MLP}}$ and $\mathcal{M}_{\mathrm{CNN}}$ see their structural and statistical performance drastically decreasing after a short time $t \approx 5$, as seen in Fig. \ref{fig:apriori_scalar_decay_stats} on the $\mathcal{L}_{\mathrm{rms}}$, correlation $\mathcal{P}$, and statistical distance $\mathcal{J}$. Interestingly, we observe that $\mathcal{M}_{\mathrm{CNN}}$ is producing a dissipation error which does not tend toward zero compared to the other models. 
Similar to the transitional regime, the data slices in Fig. \ref{fig:slices_scalar_decay} show clear evidence that the $\mathcal{M}_{\mathrm{MLP}}$ is predicting a large SGS term when the scalar is almost completely mixed. Note that the same behavior is observed for the $\mathcal{M}_{\mathrm{CNN}}$ with a smaller magnitude, but correctly tends to zero for the $\mathcal{M}_{\mathrm{SGTNN}}$ as expected from the DNS.

Overall, when comparing $\mathcal{M}_{\mathrm{CNN}}$ and $\mathcal{M}_{\mathrm{SGTNN}}$, we can say that the imposed invariances act as physical regularizers since they help the generalization while ensuring short-time coherent behavior in different \change{regimes} (first iterations in transitional regime and last iterations during scalar decay).

\subsection{A posteriori evaluation}
We now run new simulations using the NN models to compute the unknown SGS term at every iteration. These a posteriori tests are complementary to the a priori tests, because they show the behavior and time evolution of the SGS models in practice. In the following tests, we use an hybrid DNS-LES approach in order to isolate the subgrid-scalar modeling from other sources of error. To achieve this, we keep the velocity fields at DNS resolution, i.e., $512^3$, while the scalar fields of the different models evolve at the same resolution as the filtered training data, i.e., $64^3$. At every substep in the time advancement integration, the velocity fields are extrapolated from the DNS to the LES grid in spectral space. This spectral extrapolation is equivalent to a spectral cut-off filter. Note that only data on LES grid are used to obtain SGS term predictions from algebraic and NN models. The advantage of this procedure is that there is  no modeling error on the velocity field used in the scalar equation. Thus, when the scalar LES data are compared with the filtered DNS data, the difference will be due only to the scalar closure term \cite{balarac2013dynamic, vollant2017subgrid}.

The predictions provided by the different models are able to close the filtered equation \eqref{eq:filtered_advection} such that $\mathcal{M}_{\mathrm{NN}}(\overline{\mathbf{u}}, \overline{\Phi}) = \nabla \cdot \mathbf{s}$ given the low-resolution scalar and filtered velocities. Now we look at different statistical metrics such as the filtered scalar variance $\langle \overline{\Phi}^2 \rangle - \langle \overline{\Phi} \rangle^2$ and the resolved scalar enstrophy $\langle \overline{\omega}^2 \rangle / 2 = \langle \nabla^2\overline{\Phi} \rangle$, which is particularly interesting to characterizes the energy transfers in the smallest resolved scales \cite{san2015posteriori}, from which the SGS model is the most important. For each \change{regime}, we also plot the scalar energy spectrum computed using the different SGS models at intermediate and final time in the simulation.

\subsubsection{Developed turbulence regime}

\begin{figure*}[t]
  \centering
  \begin{minipage}[c]{.49\textwidth}
    \centering
    \includegraphics[width=1.\linewidth]{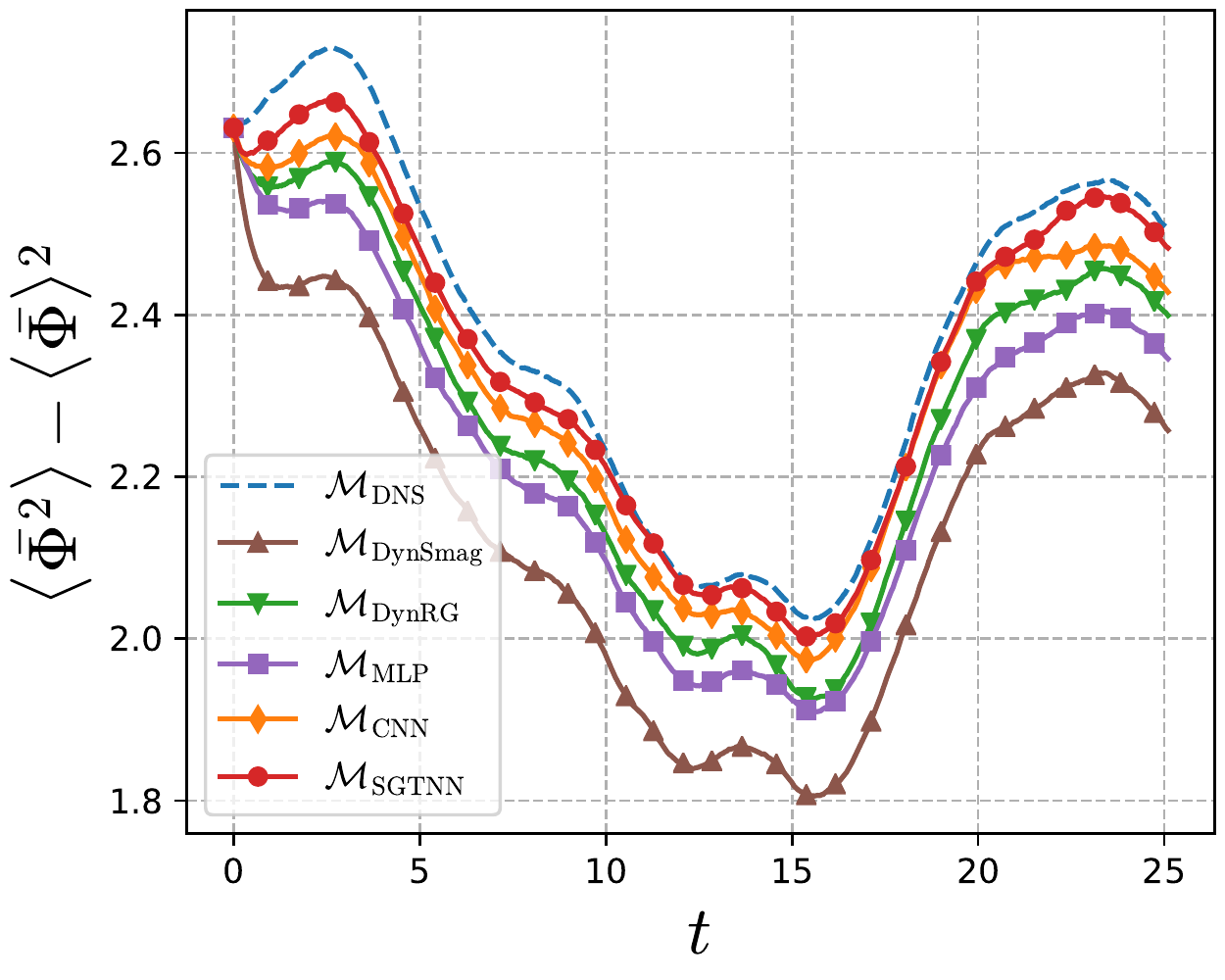}
  \end{minipage}
  \begin{minipage}[c]{.49\textwidth}
    \centering
    \includegraphics[width=1.\linewidth]{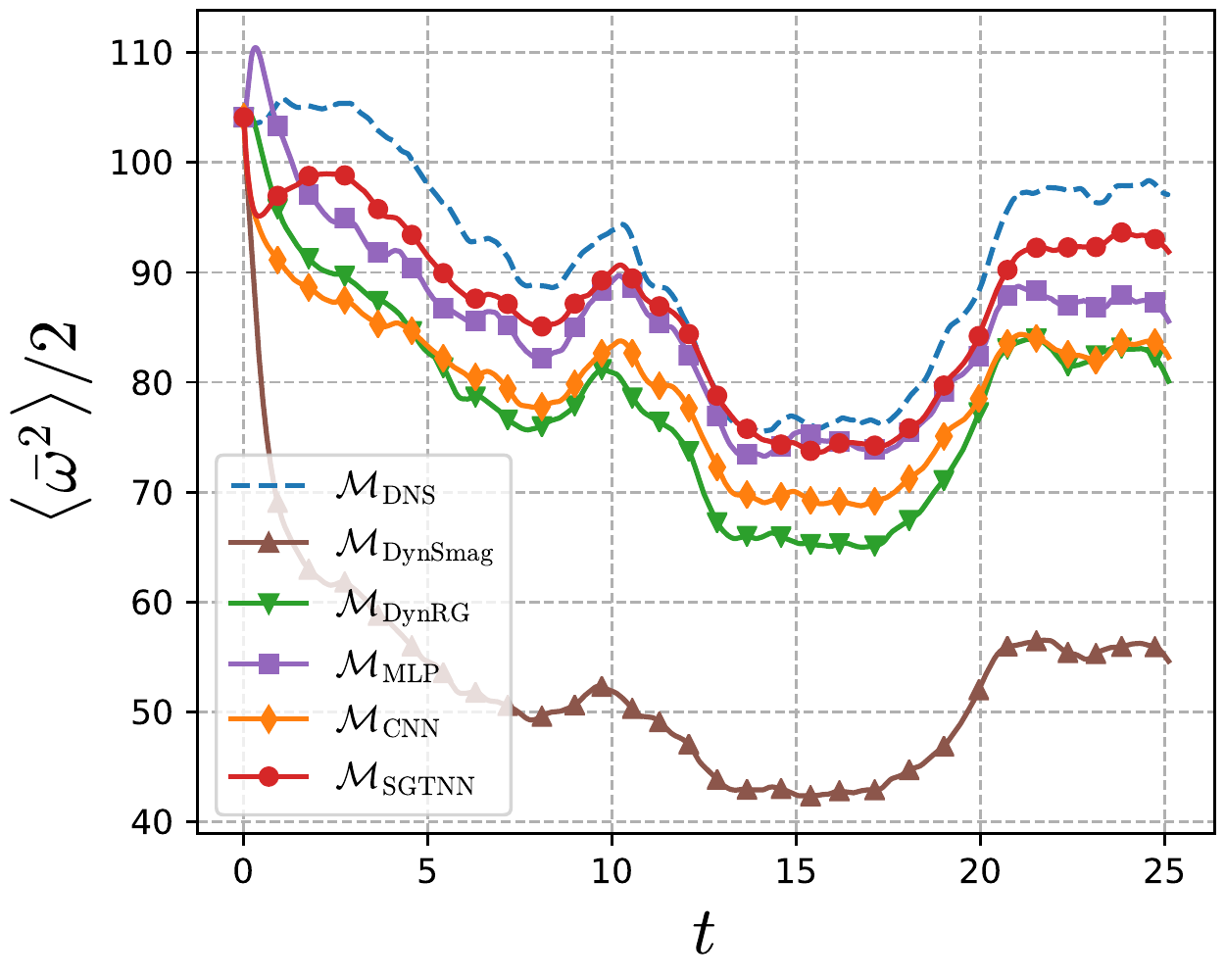}
  \end{minipage}
  \caption{Statistics of simulation in developed turbulence regime starting from training data. Scalar variance (left), resolved scalar enstrophy (right).
  \label{fig:stats_same_dev_vel}
  }
\end{figure*}

\begin{figure*}[t]
  \centering
  \begin{minipage}[c]{.49\textwidth}
    \centering
    \includegraphics[width=1.\linewidth]{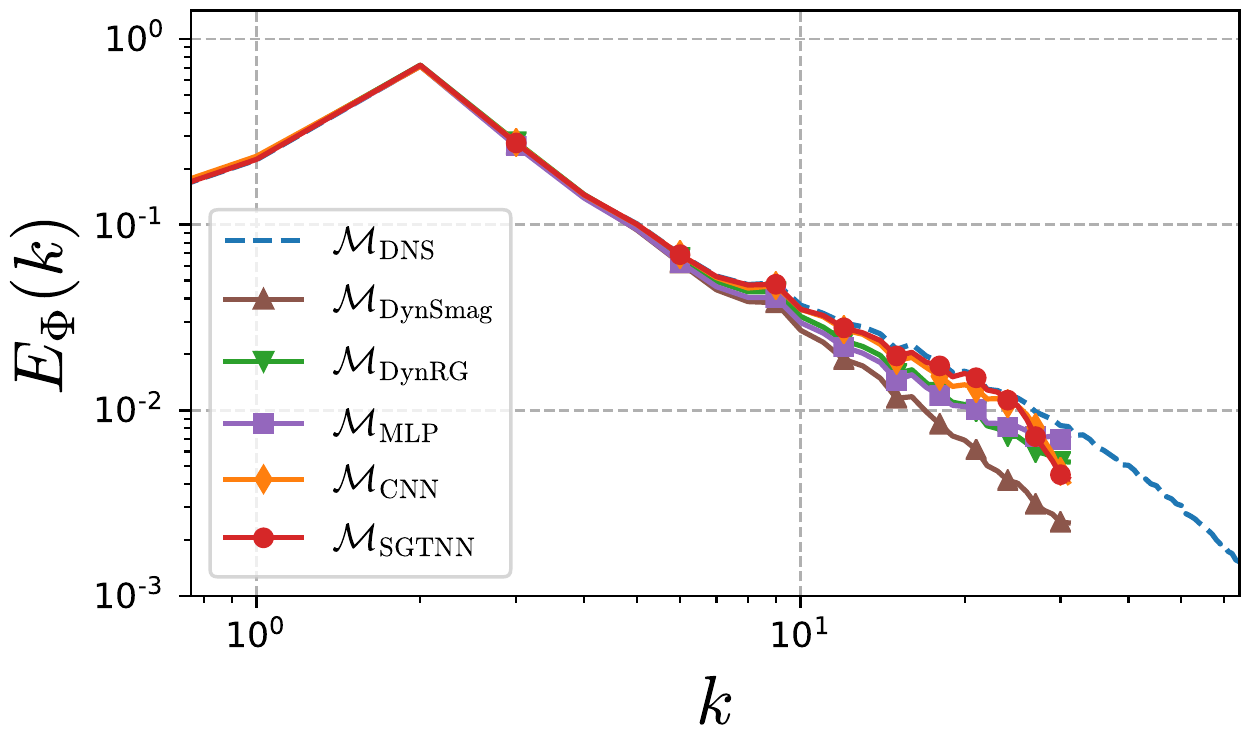}
  \end{minipage}
  \begin{minipage}[c]{.49\textwidth}
    \centering
    \includegraphics[width=1.\linewidth]{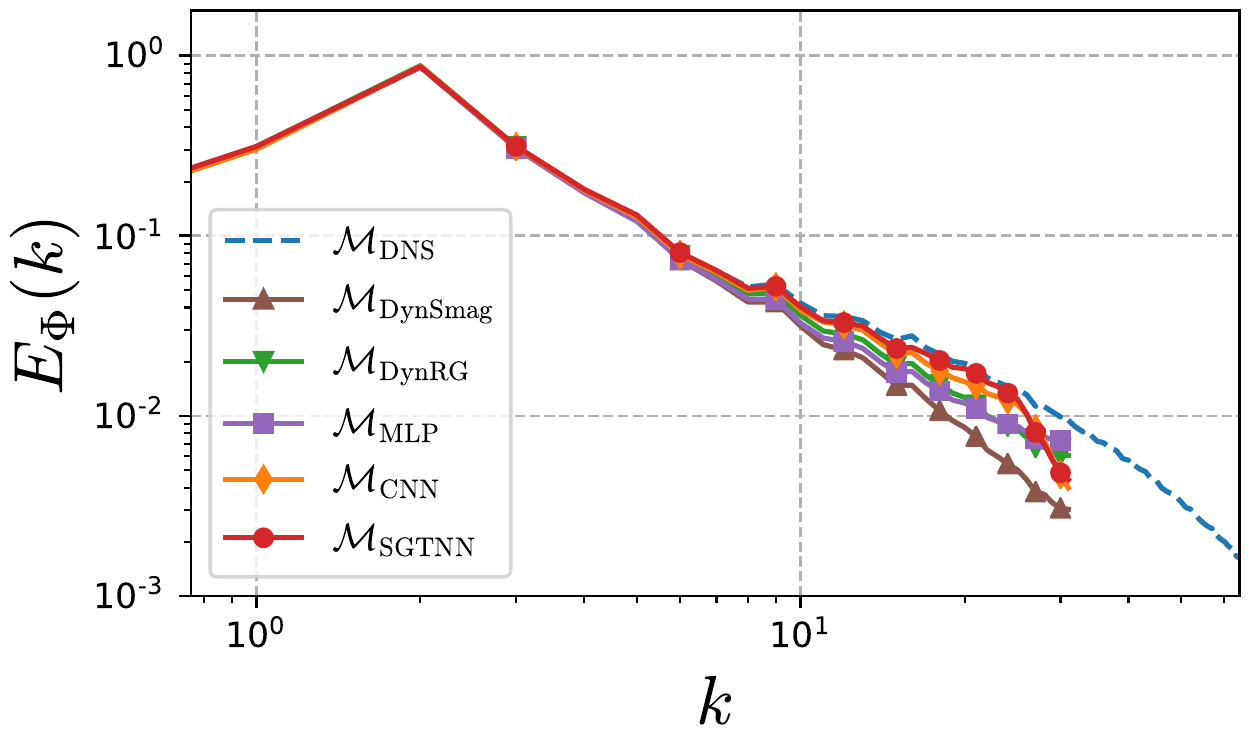}
  \end{minipage}
  \caption{Energy spectrum of simulation in developed turbulence regime starting from training data at $t = 12.5$ (left) and $t = 25$ (right).
  \label{fig:spec_same_dev_vel}
  }
\end{figure*}

\begin{figure*}[t]
  \centering
  \begin{minipage}[c]{.49\textwidth}
    \centering
    \includegraphics[width=1.\linewidth]{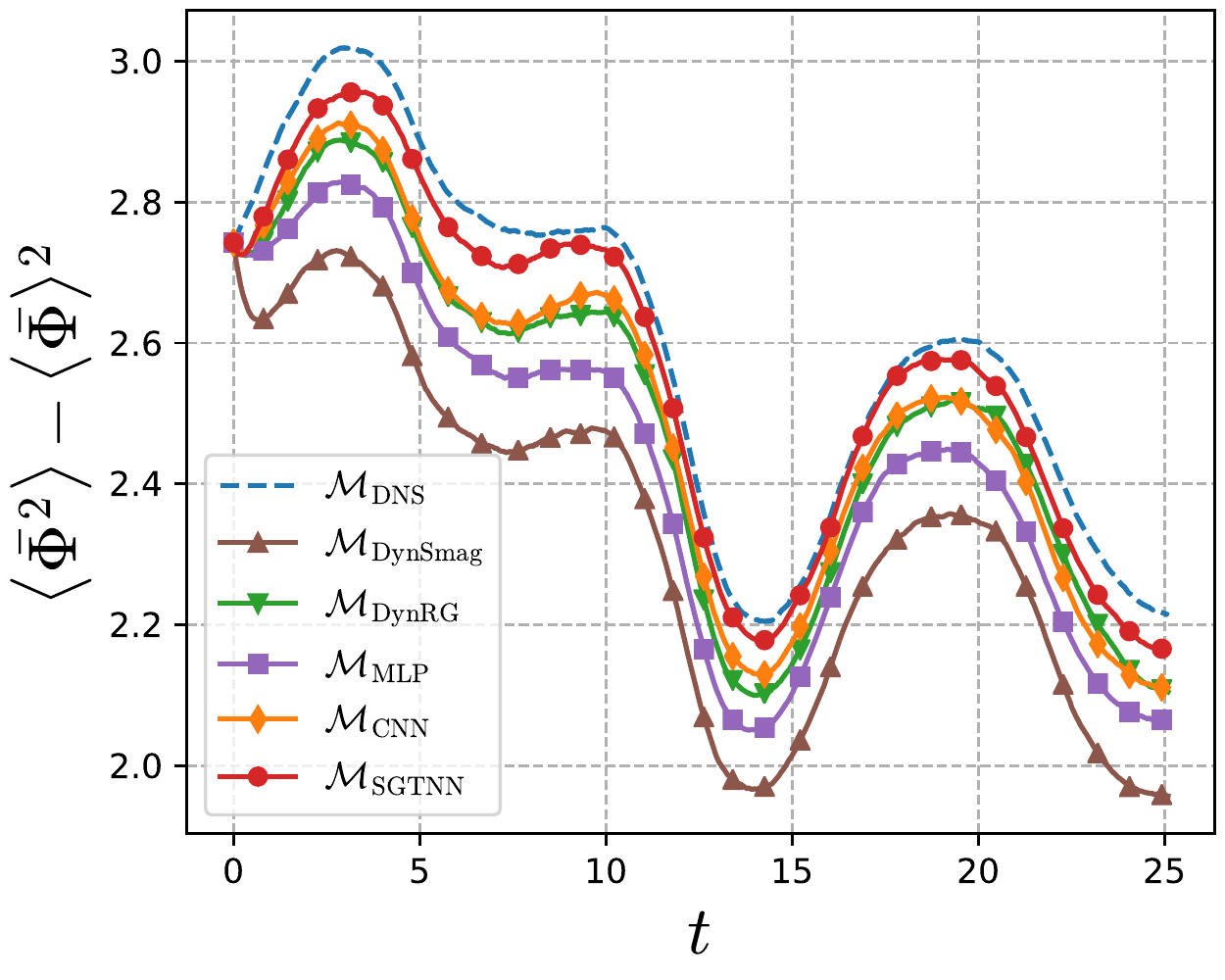}
  \end{minipage}
  \begin{minipage}[c]{.49\textwidth}
    \centering
    \includegraphics[width=1.\linewidth]{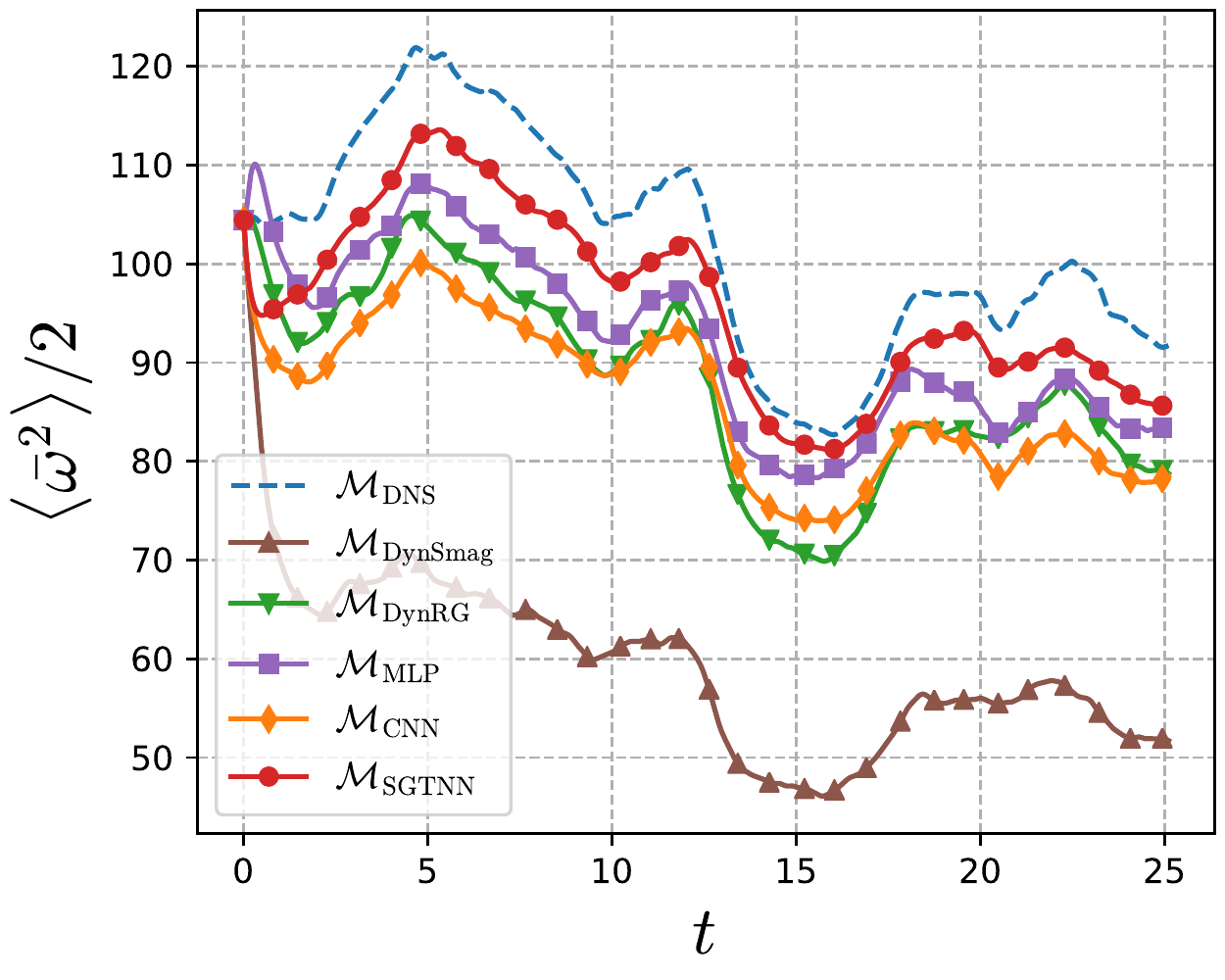}
  \end{minipage}
  \caption{Statistics of simulation in developed turbulence regime starting from testing data. Scalar variance (left), resolved scalar enstrophy (right).
  \label{fig:stats_diff_dev_vel}
  }
\end{figure*}

\begin{figure*}[t]
  \centering
  \begin{minipage}[c]{.49\textwidth}
    \centering
    \includegraphics[width=1.\linewidth]{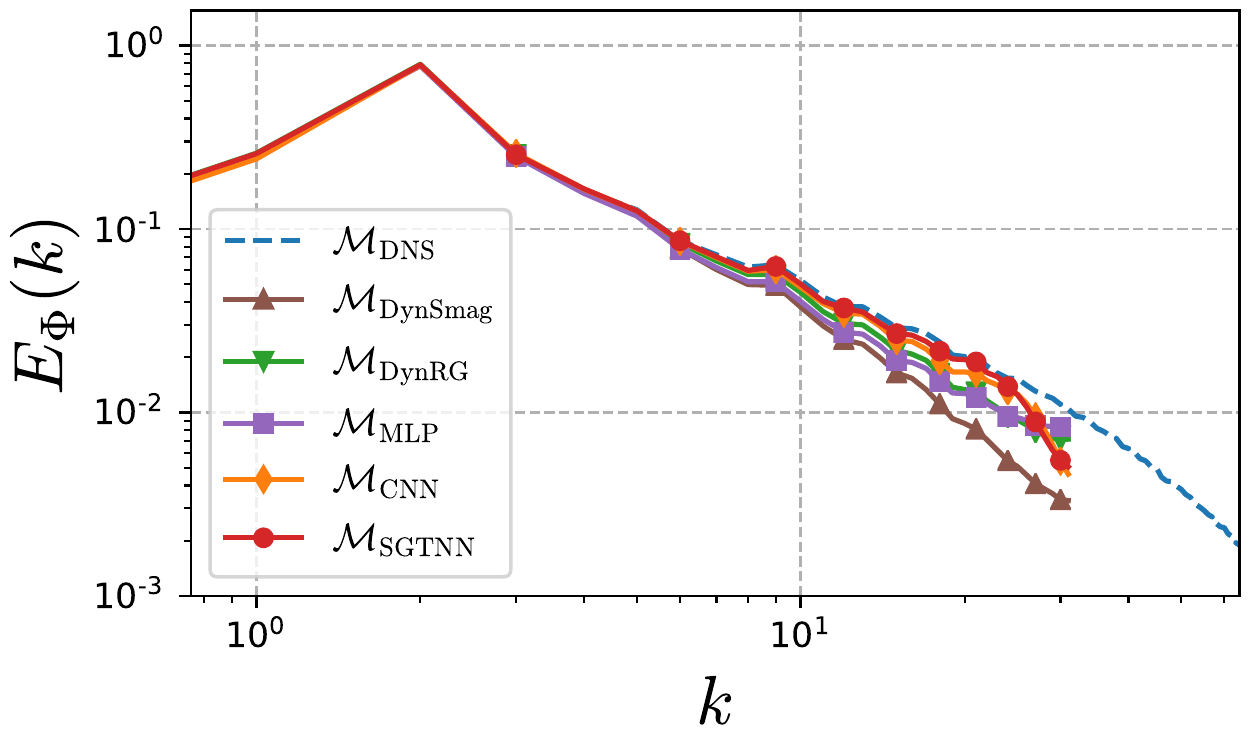}
  \end{minipage}
  \begin{minipage}[c]{.49\textwidth}
    \centering
    \includegraphics[width=1.\linewidth]{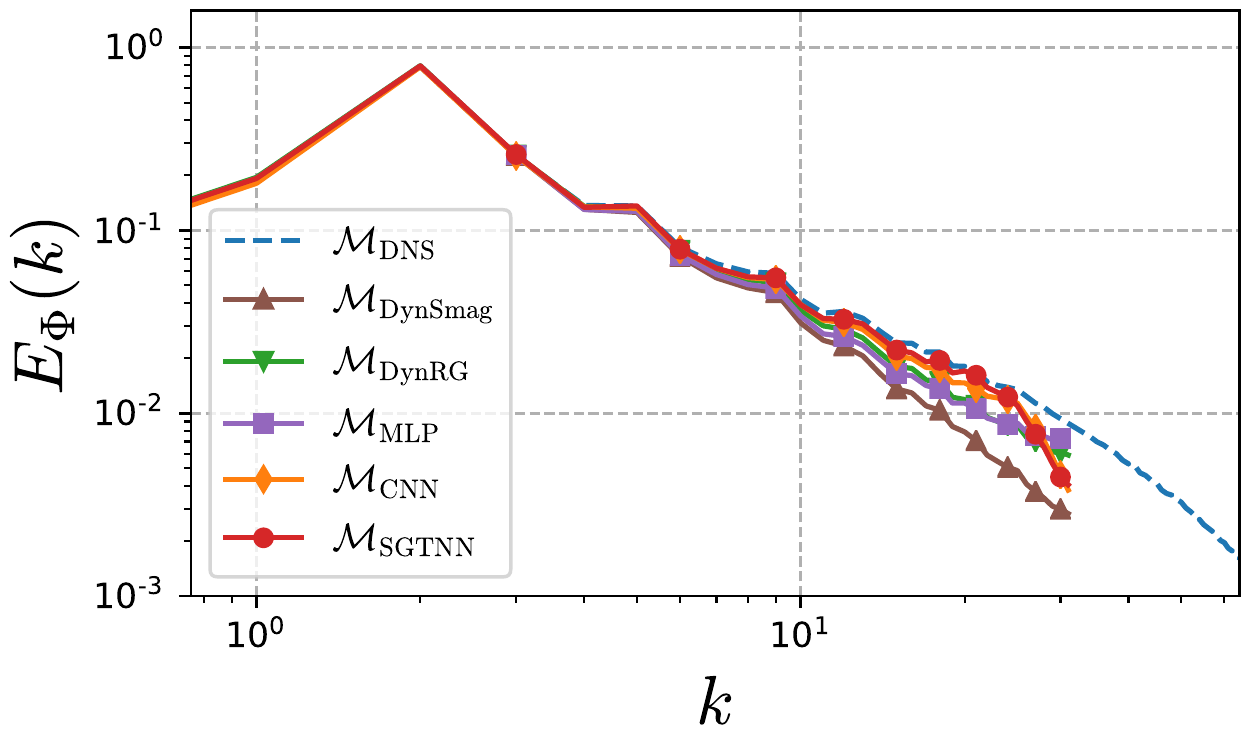}
  \end{minipage}
  \caption{Energy spectrum of simulation in developed turbulence regime starting from testing data at $t = 12.5$ (left) and $t = 25$ (right).
  \label{fig:spec_diff_dev_vel}
  }
\end{figure*}

In the first two simulations, we start in a developed turbulence regime at $\mathcal{S} = 20000$ of the training data (see Fig. \ref{fig:train_data}) and the testing data (see Fig. \ref{fig:tests_data}), respectively, with the same velocity and scalar forcing as used in the datasets. Using the same starting point as the training data is useful in order to consider the accumulation of errors during the simulation, since the models learnt from only a finite number of samples of the exact simulation. We see in Fig. \ref{fig:stats_same_dev_vel} that the scalar variance, on the left, produced by the $\mathcal{M}_{\mathrm{SGTNN}}$ and the $\mathcal{M}_{\mathrm{CNN}}$ are the closest from the DNS except after a long integration time, $t \approx 20$ where the $\mathcal{M}_{\mathrm{CNN}}$ starts accumulating error. The behavior of the model on the largest scales in shown by the resolved scalar enstrophy on the right. Here, the $\mathcal{M}_{\mathrm{MLP}}$ has a consistent behavior almost similar to the $\mathcal{M}_{\mathrm{SGTNN}}$. One can verify in the spectra from Fig. \ref{fig:spec_same_dev_vel} that the two models presented in this work are the most accurate on the smallest wave-numbers, while the NN baseline model from Ref. \cite{portwood2021interpreting} is most effective close to the spectral cut at $k \geq 30$. Now, the same conclusions can be drawn with the testing data starting point as depicted in Figs. \ref{fig:stats_diff_dev_vel} and \ref{fig:spec_diff_dev_vel}. We can already see that a CNN without physical constraints is not able to reproduce the dynamics of the SGS term in statistically similar conditions as those of the training data. Indeed, it performs at the same accuracy for the scalar variance but worse for the resolved scalar enstrophy than some algebraic models. The effect of the physical invariances is primordial and gives the ability to get the best performances from the neural network.

\subsubsection{Scalar transition regime}

\begin{figure*}[t]
  \centering
  \begin{minipage}[c]{.49\textwidth}
    \centering
    \includegraphics[width=1.\linewidth]{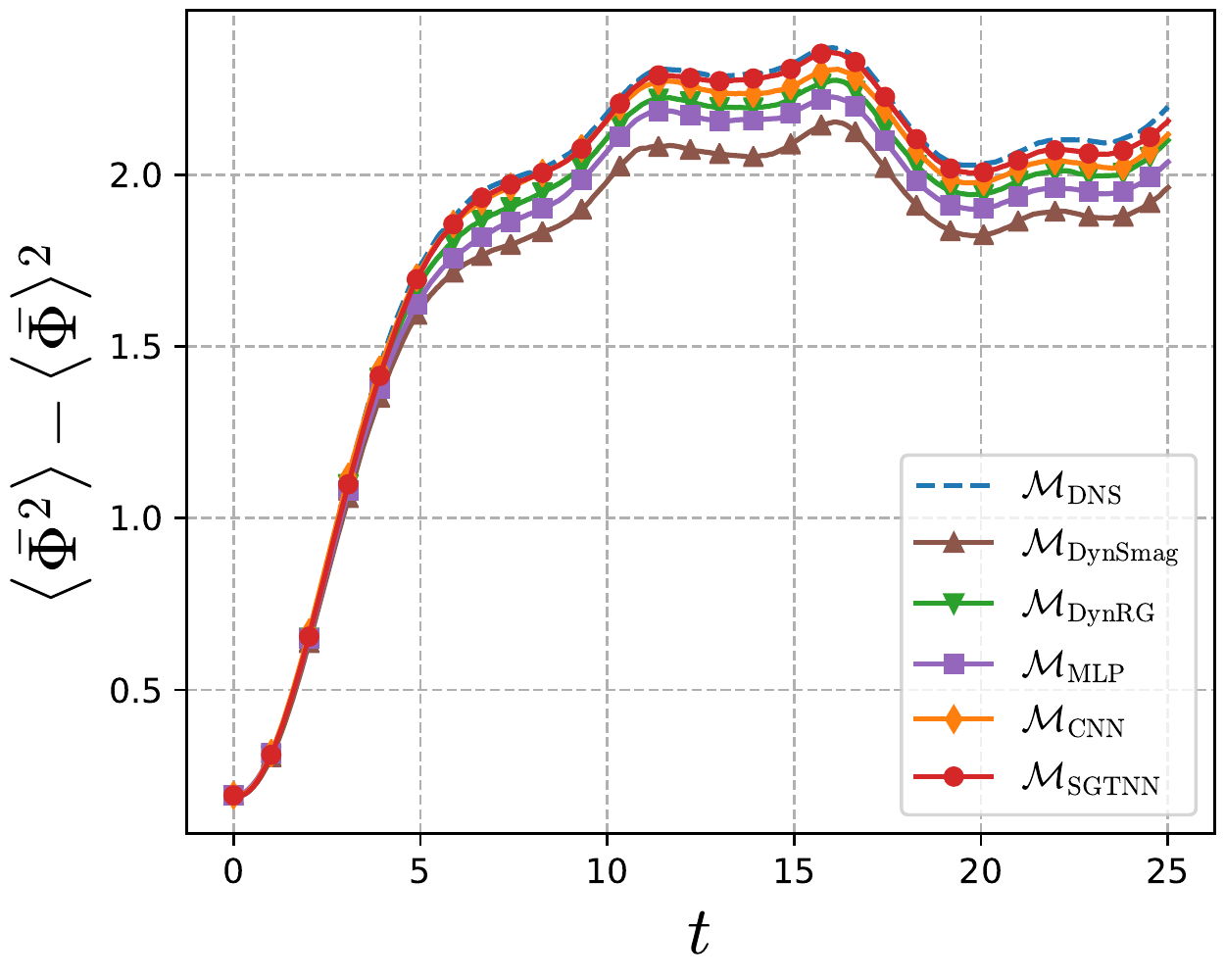}
  \end{minipage}
  \begin{minipage}[c]{.49\textwidth}
    \centering
    \includegraphics[width=1.\linewidth]{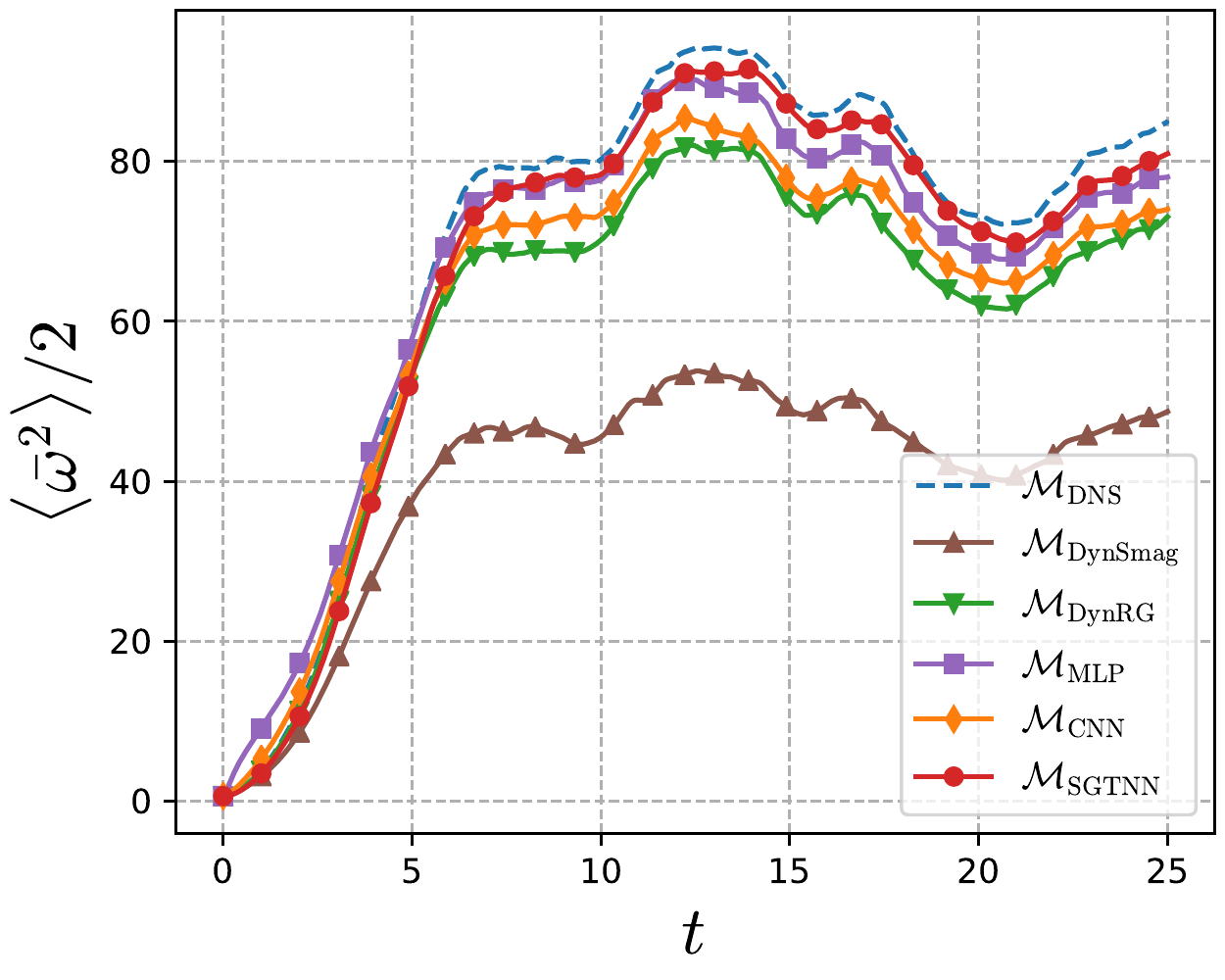}
  \end{minipage}
  \caption{Statistics of simulation in scalar transition regime, starting from a newly initialized field. Scalar variance (left), resolved scalar enstrophy (right).
  \label{fig:stats_diff_scalar_transition}
  }
\end{figure*}

\begin{figure*}[t]
  \centering
  \begin{minipage}[c]{.49\textwidth}
    \centering
    \includegraphics[width=1.\linewidth]{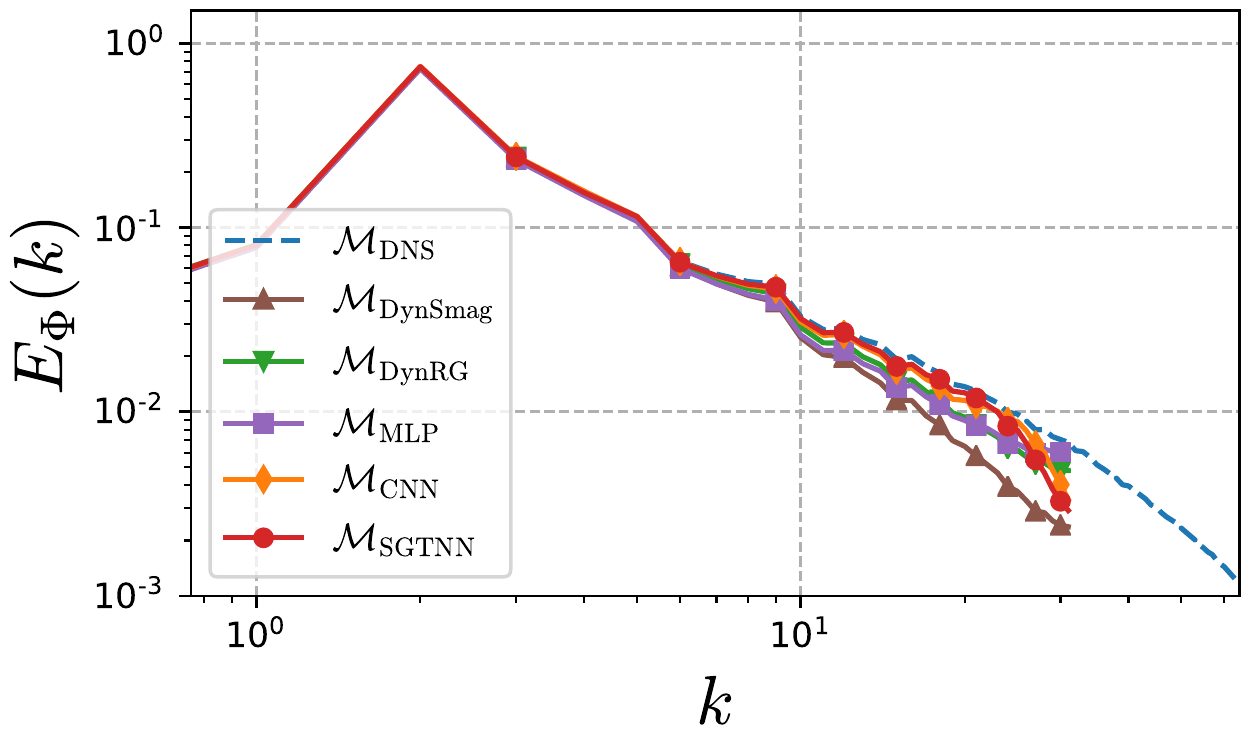}
  \end{minipage}
  \begin{minipage}[c]{.49\textwidth}
    \centering
    \includegraphics[width=1.\linewidth]{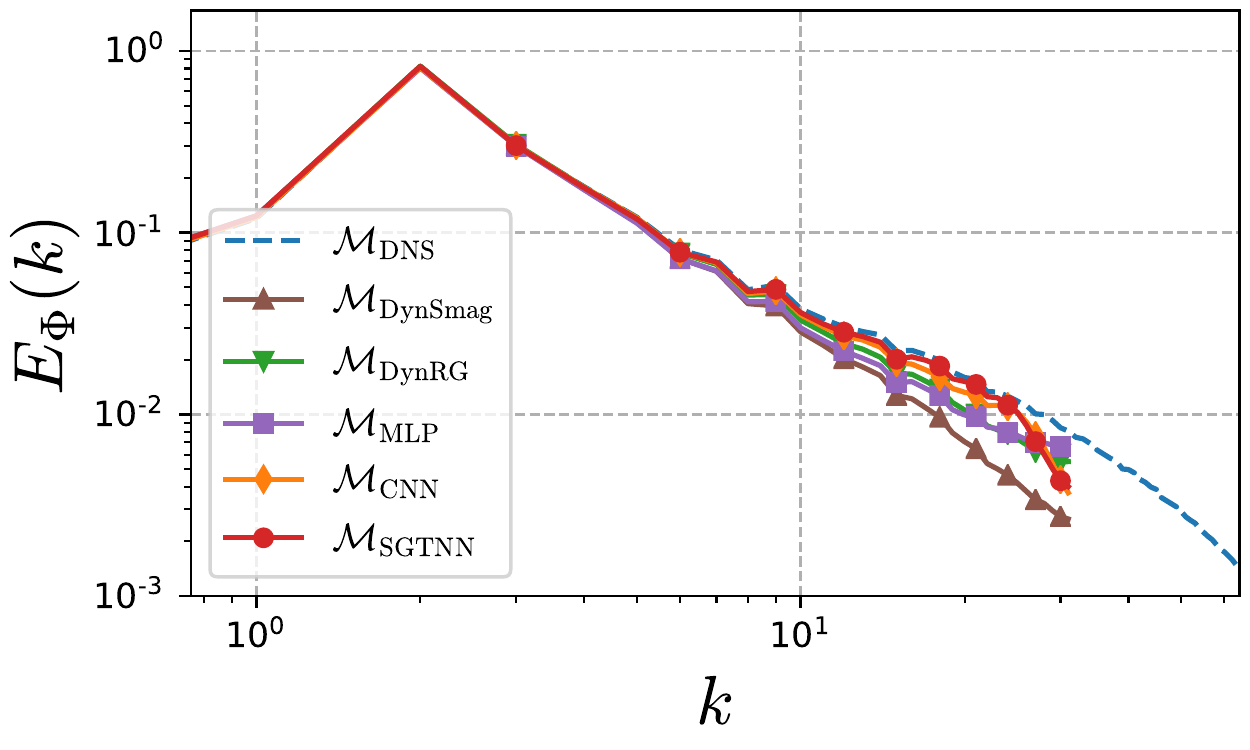}
  \end{minipage}
  \caption{Energy spectrum of simulation in scalar transition regime, starting from a newly initialized field at $t = 6$ (left) and $t = 25$ (right).
  \label{fig:spec_diff_scalar_transition}
  }
\end{figure*}

The scalar transition regime is already testing the extrapolation capabilities of the model. In this \change{regime}, we start from a velocity field already in turbulent motion and a scalar initialized at large scales and forced with the previously used scheme. The difficulty of this \change{regime} resides in the transition of the scalar field to turbulent advection, which then comes down to the previous \change{regime} after $t = 15$. Both scalar variance and resolved enstrophy are very well reproduced by the $\mathcal{M}_{\mathrm{SGTNN}}$ (see Fig. \ref{fig:stats_diff_scalar_transition}), which gives superior performances compared to the algebraic models. The hypothesis drawn from the a priori test of the similar \change{regime} seems to correlate with the a posteriori statistics, where both $\mathcal{M}_{\mathrm{CNN}}$ and $\mathcal{M}_{\mathrm{MLP}}$ produce a large error during the first transitional iterations, while the invariant NN model is more consistent. However, Fig. \ref{fig:spec_diff_scalar_transition} indicates that the spectra produced by the CNN models are closer to those of the DNS compared to the baseline models. In particular, we can see on the left that most significant deviation appears during the transition to turbulent advection, i.e., the first iterations of the simulation.

\subsubsection{Full decay regime}

\begin{figure*}[t]
  \centering
  \begin{minipage}[c]{.49\textwidth}
    \centering
    \includegraphics[width=1.\linewidth]{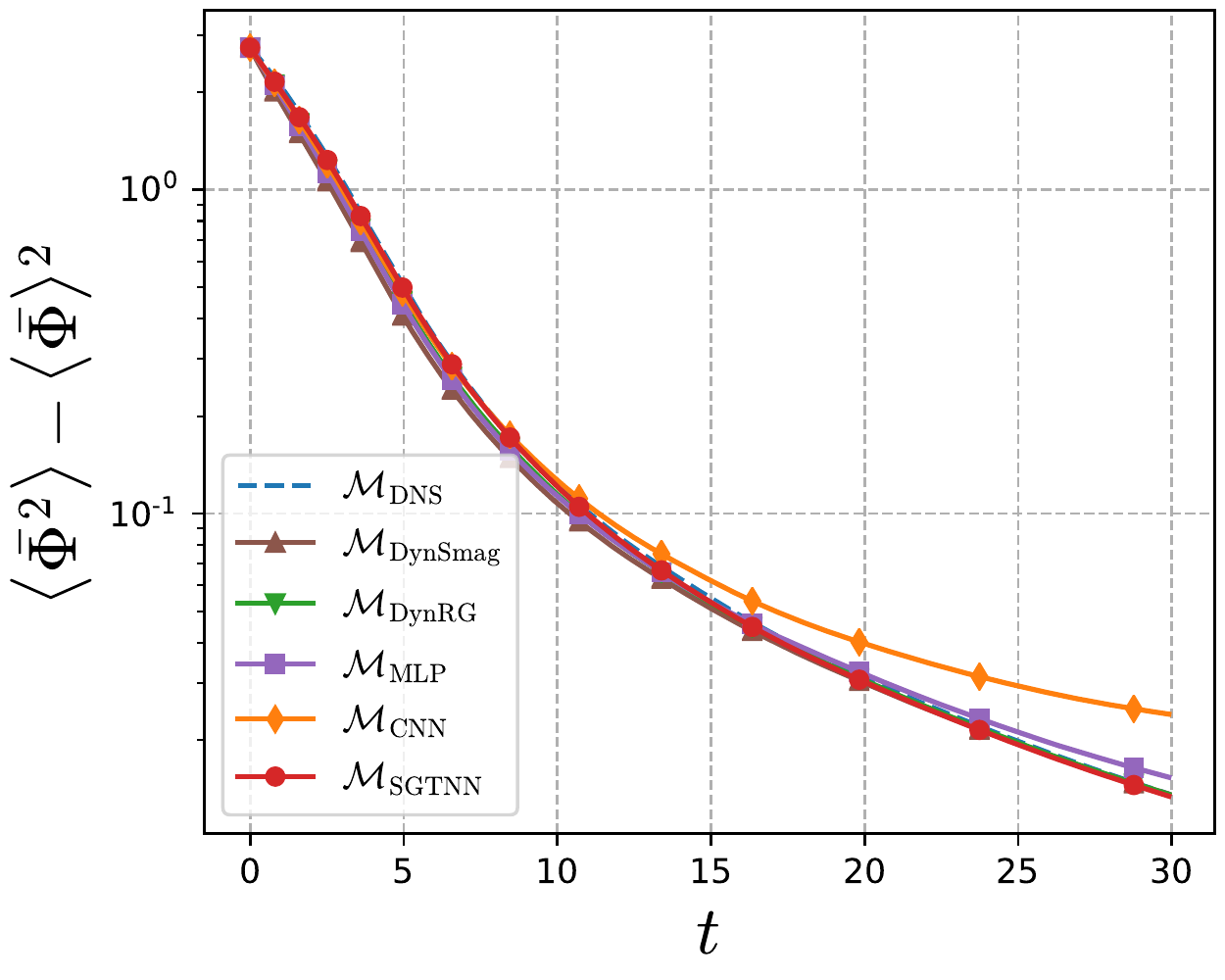}
  \end{minipage}
  \begin{minipage}[c]{.49\textwidth}
    \centering
    \includegraphics[width=1.\linewidth]{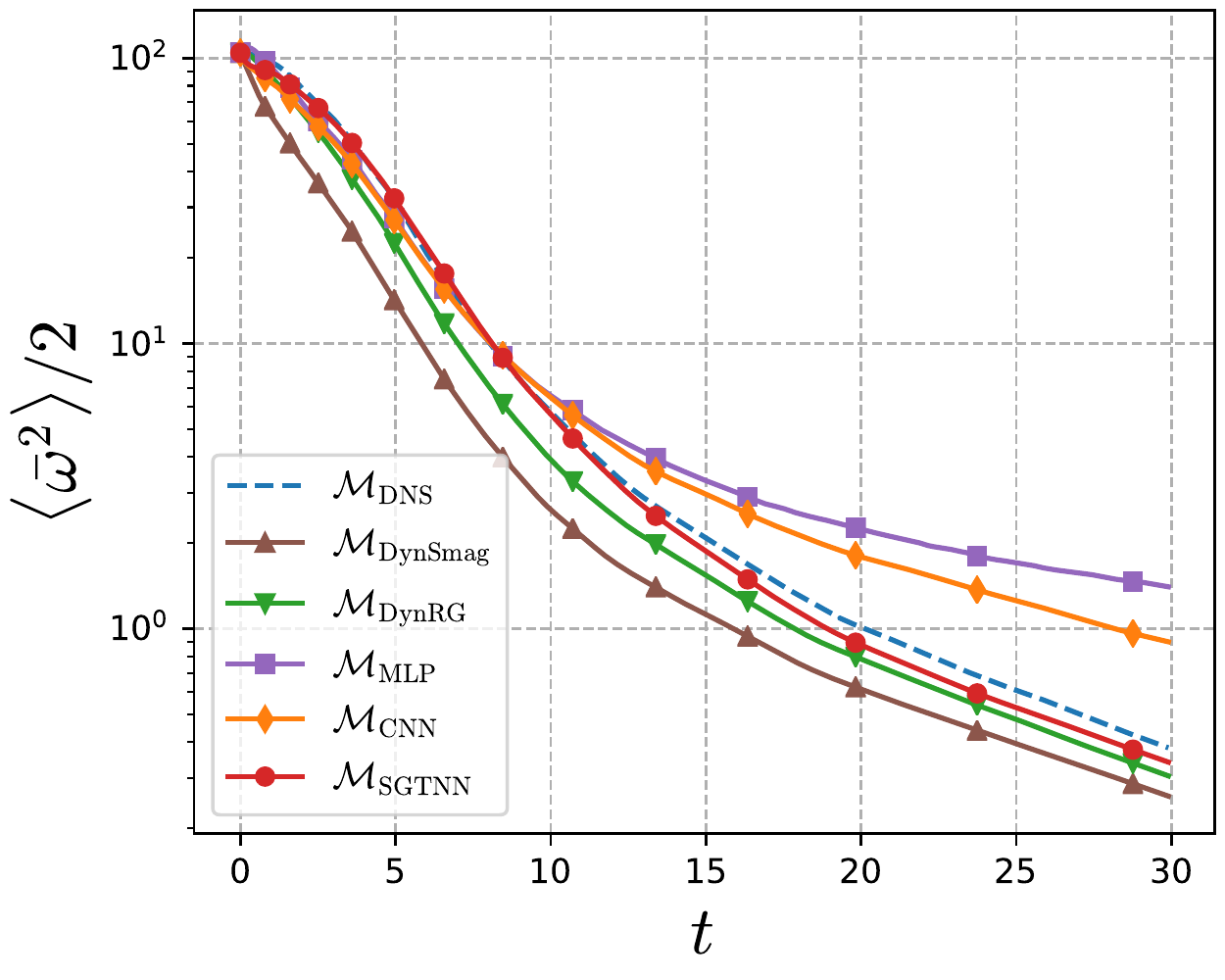}
  \end{minipage}
  \caption{Statistics of simulation in full decay regime starting from testing data. Scalar variance (left), resolved scalar enstrophy (right). 
  \label{fig:stats_diff_full_decay}
  }
\end{figure*}
   
\begin{figure*}[t]
  \centering
  \begin{minipage}[c]{.49\textwidth}
    \centering
    \includegraphics[width=1.\linewidth]{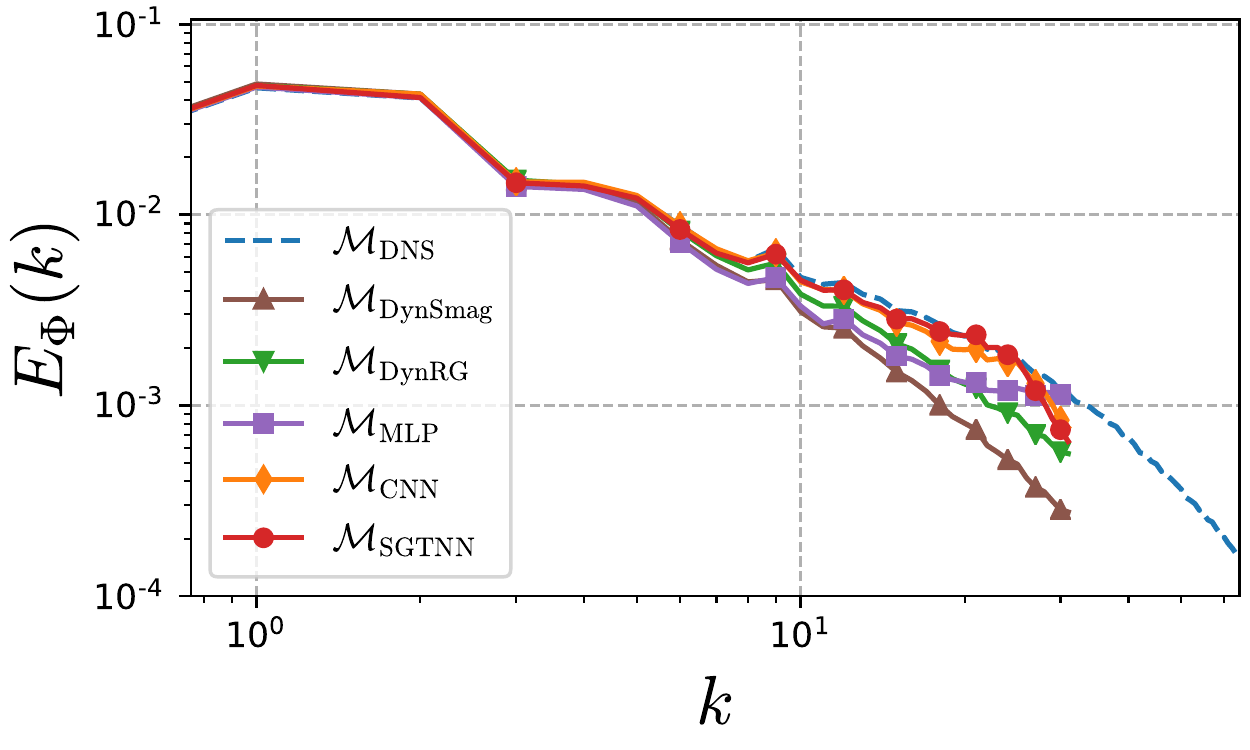}
  \end{minipage}
  \begin{minipage}[c]{.49\textwidth}
    \centering
    \includegraphics[width=1.\linewidth]{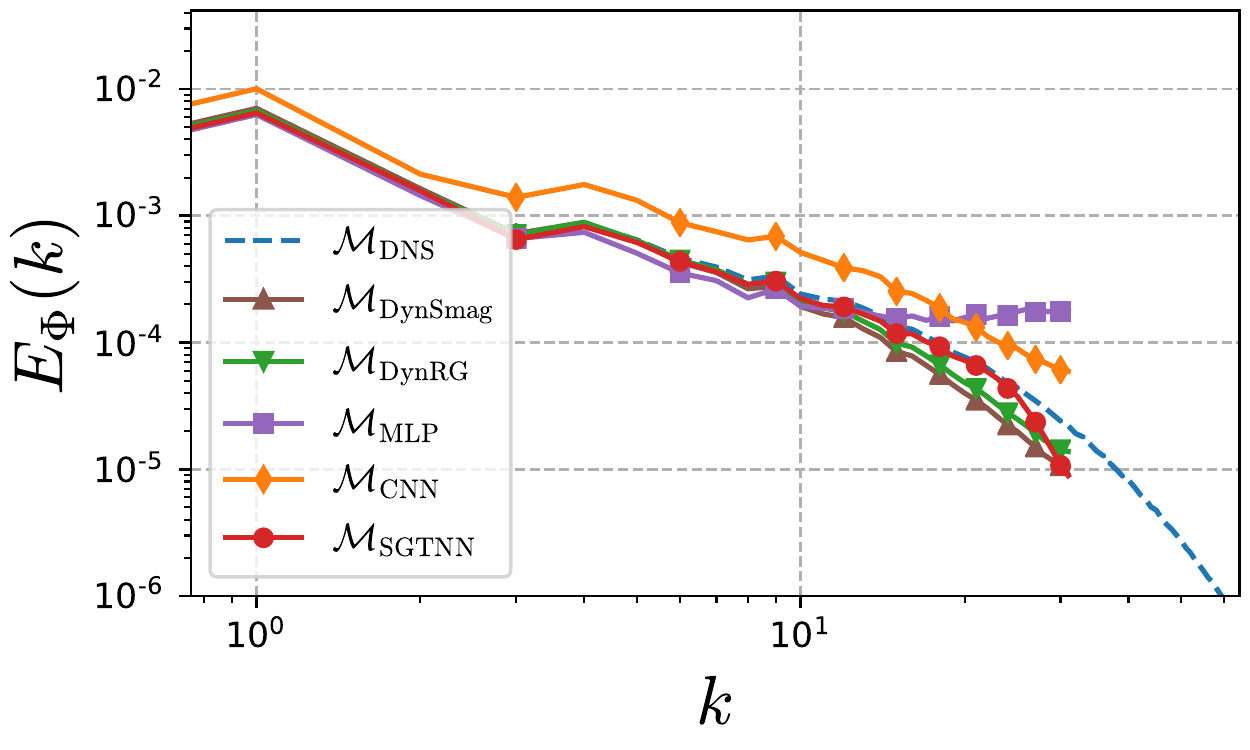}
  \end{minipage}
  \caption{Energy spectrum of simulation in full decay regime starting from testing data at $t = 8$ (left) and $t = 30$ (right).
  \label{fig:spec_diff_full_decay}
  }
\end{figure*}

We saw in a priori tests that the $\mathcal{M}_{\mathrm{SGTNN}}$ gave the best performances in the scalar decay \change{regime} while the other NN models were not able to generalize. In a posteriori, we go one step further and completely remove the source terms for both velocity and scalar fields, which results in a full decay. However, we still observe the same behavior in Fig. \ref{fig:stats_diff_full_decay}. More precisely, the resolved scalar enstrophy shows that the $\mathcal{M}_{\mathrm{MLP}}$ and $\mathcal{M}_{\mathrm{CNN}}$ are still producing energy transfers after a long time. We also see in the spectrum (Fig. \ref{fig:spec_diff_full_decay}) that both unconstrained NN models are not able to dissipate the large scales of the SGS term at $t = 8$, which results in a blowup of the simulation after $t = 30$. The $\mathcal{M}_{\mathrm{SGTNN}}$, however, is still in excellent agreement with the DNS and out-performs significantly the algebraic models, even in \change{regimes} that were not part of the training data.

\subsubsection{Generalization capabilities}
The a posteriori \change{regimes} discussed above are statistically different from the data used to train the NN models. To evaluate the robustness of the models and their ability to extrapolate to different \change{regimes}, we compute the largest absolute error of the previously shown flow statistics, namely, scalar variance and enstrophy during their time evolution,
\begin{equation}
    \mathcal{L}_{\mathrm{max}}(X, Y) = \max |X - Y| / Y,
\end{equation}
where $Y$ and $X$ refer to the flow statistics computed from DNS simulations and from LES simulations with the SGS model, respectively.
The results are shown in Table. \ref{tbl:aposteriori_gen} for the scalar variance and resolved scalar enstrophy. For each \change{regime}, we also compare the current $\mathcal{L}_{\mathrm{max}}(X, Y)$ with the reference (or base) given by the developed turbulence \change{regime} starting from training data. For the scalar variance, the results given by the $\mathcal{M}_{\mathrm{MLP}}$ are quite robust in the different \change{regimes}, and the maximum error in full decay is only $\times 1.8275$ higher compared to the base \change{regime}. The $\mathcal{M}_{\mathrm{CNN}}$ is consistent in the first \change{regimes}, but its performance dramatically drops in the full decay. The $\mathcal{M}_{\mathrm{SGTNN}}$, however, is able to maintain a stable performance with an error in full decay only $\times 1.1509$ higher than in developed turbulence. For the resolved scalar enstrophy, we can see that the $\mathcal{M}_{\mathrm{MLP}}$ is not even robust to the scalar transition, with an error more than $\times 20$ higher than in the base \change{regime}. The $\mathcal{M}_{\mathrm{CNN}}$ sees its performance gradually degraded throughout the \change{regimes} of increasing difficulty. The same comments are also valid for the $\mathcal{M}_{\mathrm{SGTNN}}$, except that the maximum error has increased by a factor of $\times 1.6841$ in full decay, which is five times smaller than the error increase of the model without physical invariances.

\begin{table*}[t]
\begin{ruledtabular}
\begin{tabular}{lcccc}
& $\mathcal{L}_{\mathrm{max}}$ (base) & $\mathcal{L}_{\mathrm{max}}$ & $\mathcal{L}_{\mathrm{max}}$ & $\mathcal{L}_{\mathrm{max}}$\\
$X = \langle \bar{\Phi}^{2} \rangle - \langle \bar{\Phi} \rangle^{2}$ &&&&\\
\hline
$\mathcal{M}_{\mathrm{MLP}}$   & 0.0719 ($\times$1) & 0.0839 ($\times$1.1669) & 0.0731 ($\times$1.0167) & 0.1314 ($\times$1.8275)\\
$\mathcal{M}_{\mathrm{CNN}}$   & 0.0405 ($\times$1) & 0.0578 ($\times$1.4272) & 0.0389 ($\times$0.9605) & 0.7650 ($\times$18.8889)\\
$\mathcal{M}_{\mathrm{SGTNN}}$ & \textbf{0.0265} ($\times$1) & \textbf{0.0268} (\textbf{$\times$1.0113}) & \textbf{0.0193} (\textbf{$\times$0.7283}) & \textbf{0.0305} (\textbf{$\times$1.1509})\\
\hline
$X = \langle \bar{\omega}^{2} \rangle / 2$ &&&&\\
\hline
$\mathcal{M}_{\mathrm{MLP}}$   & 0.1180 ($\times$1) & 0.1271 (\textbf{$\times$1.0771}) & 2.6277 ($\times$22.2686) & 2.6676 ($\times$22.6068)\\
$\mathcal{M}_{\mathrm{CNN}}$   & 0.1747 ($\times$1) & 0.1890 ($\times$1.0818) & 0.4938 ($\times$2.8266)  & 1.3529 ($\times$7.7441)\\
$\mathcal{M}_{\mathrm{SGTNN}}$ & \textbf{0.0823} ($\times$1) & \textbf{0.0908} ($\times$1.1033) & \textbf{0.1117} (\textbf{$\times$1.3572}) & \textbf{0.1386} (\textbf{$\times$1.6841})\\
\hline
& Developed (train) & Developed (test) & Scalar transition & Full decay
\end{tabular}
\end{ruledtabular}
\caption{Maximum error $\mathcal{L}_{\mathrm{max}}$ on scalar variance (first three rows) and resolved scalar enstrophy (last three rows) of the different NN models during each a posteriori \change{regime}. Relative error compared to the base \change{regime}, i.e., developed turbulence starting from training data is shown as $\mathcal{L}_{\mathrm{max}} /$ base. \label{tbl:aposteriori_gen}}
\end{table*}

\section{Summary}
We propose a closure model to the problem of scalar transport in a turbulent incompressible flow based on NNs and physical invariances. The model is shown to predict a realistic behavior and out-performs classical algebraic models on different metrics. During the a posteriori evaluation, we demonstrate the capabilities of the model to generalize to unseen flow \change{regimes} compared to NN models that do not embed physical knowledge. Thus, we think that to be applicable to large scale simulations, NN models will require some form of physical invariance. Other interesting directions towards a generic model would include an explicit mention of the filter width, Reynolds number and Schmidt number. Learning a dynamic coefficient related to the filter width could be the subject of future investigations. The ideal setting of spectral discretization also introduces some limitations with respect to the geometry of the domain. Moving to classical numerical methods such as finite volumes or finite differences constitutes a next exploratory step that comes with difficulties related to the separation of physical quantities and discretization errors introduced by the chosen scheme.

\section{Acknowledgements}
This work was supported by the CNRS through the 80 PRIME project and the ANR through the Melody, OceaniX, and HRMES projects. 
Computations were performed using HPC and GPU resources from GENCI-IDRIS (Grants 020611 and 101030) and GRICAD infrastructures.

\bibliography{ref}

\end{document}